\documentclass[11pt,draftclsnofoot,journal,onecolumn]{IEEEtran}

\IEEEoverridecommandlockouts

\usepackage{graphicx}
\usepackage{amsmath}
\usepackage{amsfonts}
\usepackage{amssymb}
\usepackage{array}
\usepackage{cite}
\usepackage{url}
\usepackage{psfrag}
\usepackage{subfigure}
\usepackage[ruled,vlined,shortend]{algorithm2e}
\usepackage{algorithmic}

\newtheorem{proposition}{Proposition}

\renewcommand{\IEEEQED}{\IEEEQEDopen}

\begin{document}

\title{Practical Rate and Route Adaptation with Efficient Link Quality Estimation for IEEE 802.11b/g Multi-Hop Networks}
\author{Jinglong {Zhou}, Vijay S. {Rao}, Przemys{\l}aw Pawe{\l}czak, Daniel {Wu}, and Prasant {Mohapatra}
\thanks{Jinglong {Zhou} and Vijay S. {Rao} are with the Department of Electrical Engineering, Mathematics and Computer Science, Delft University of Technology, Mekelweg 4, 2628 CD Delft, the Netherlands (email: \{j.l.zhou, v.sathyanarayanarao\}@tudelft.nl).}
\thanks{Przemys{\l}aw {Pawe{\l}czak} is with the Department of Electrical Engineering, University of California, Los Angeles, 56-125B Engineering IV Building, Los Angeles, CA 90095-1594, USA (email: przemek@ee.ucla.edu).}
\thanks{Daniel {Wu} and Prasant {Mohapatra} are with the Department of Computer Science, University of California, Davis, 2063 Kemper Hall, 1 Shields Avenue, Davis, CA 95616, USA (email: \{danwu, pmohapatra\}@ucdavis.edu).}
\thanks{Initial results on SNR maps-based IEEE 802.11 b/g link quality estimation presented in this paper were published in proceedings of PerNets (IEEE/Create-Net MobiQuitous 2007 workshop), Aug. 10, 2007, Philadelphia, PA, USA~\cite{Zhou_Pernets_2007} and IEEE VTC2009-Spring conference, Barcelona, Spain, EU, Apr. 26--29, 2009~\cite{Zhou_VTC_2009}.}
\thanks{Also available at http://arxiv.org/abs/0909.5263.}
}

\maketitle

\begin{abstract}
Accurate and fast packet delivery ratio (PDR) estimation, used in evaluating wireless link quality, is a prerequisite to increase the performance of multi-hop and multi-rate IEEE 802.11 wireless networks. Unfortunately, contemporary PDR estimation methods, i.e. beacon-based packet counting in Estimated Transmission Time and Expected Transmission Count metrics, have unsatisfactory performance. Therefore, in this paper we propose a novel PDR estimation method based on SNR profiles and special broadcast packets. We classify all possible link quality estimation (LQE) sources and compare them analytically against our design, showing the superiority of our approach. Further investigations with the prototype implementation of our method in IEEE 802.11b/g testbeds reveal that the LQE accuracy in some scenarios can be improved up to 50\% in comparison to generic beacon packet-based LQE. The proposed method is used in novel rate adaptation process, while these two methods are jointly used in the routing process. Experiments in different measurement scenarios show that the advanced LQE leads to a better rate adaptation and route selection in the form of end-to-end throughput increase, compared to traditional LQE methods.
\end{abstract}

\IEEEpeerreviewmaketitle

\section{Introduction}
\label{sec:introduction}

Due to the broadcast nature of the wireless medium in IEEE 802.11b/g-based wireless mesh and ad hoc networks, each node may form many links with its neighbors in a large network. The links are formed based on the data rates and data packet sizes used for neighbor discovery. To achieve high connectivity and end-to-end throughput, nodes must adapt the link data rates and routes, respectively. Wireless link quality, usually measured by Packet Delivery Ratio (PDR), is the fundamental metric by which data rate and route selection are optimized~\cite{Bicket_SampleRate_2005,BicketMobicom2005}. Therefore, efficient and accurate Link Quality Estimation (LQE) is pivotal to the effective operation of wireless networks. However, there are multiple factors that affect PDR, predominantly SNR, shadowing, multipath, interference, and packet collisions, which make effective LQE challenging. Previous results indicate that link quality can differ significantly with different propagation environments, e.g. indoor and outdoor~\cite{zuniga_2004_secon}. Unfortunately, our investigation shows that all known LQE methods do not meet the required objectives: (i) to  measure link quality efficiently, (ii) to adapt measurement result to a new environment, and (iii) maintain certain level of accurate throughout.

The popular LQE methods are based on received packet count. Usually beacon~\cite{BicketMobicom2005} or unicast~\cite{Kim2006,Kim_Ton_2009} packets are sent periodically over each link. Later, transmitters count the number of acknowledged packets from each receiver. Due to the difference between data packets and probing packets\footnote{By probing packets we mean a packet that has been sent separately from regular data packet to assess link quality.}, the neighbor discovery is inefficient and the LQE is inaccurate. Among the newer LQE methods, a promising approach is based on correlating SNR levels with PDRs, as proposed independently in~\cite{Zhou_VTC_2009,verma_2008_Wimesh,Zhou_Pernets_2007}. However the evaluation of this method is very limited and its potential unknown.

The issue of LQE adaptation to a new network environment is not well discussed in the literature either. Simply put, most of the LQE methods we are aware of are evaluated using limited network scenarios, e.g. with limited transmission rates, and in limited propagation environments~\cite{souryal_2006_wcnc,wang_2006_realman}. The impact of the rate and route selections by different LQE methods are unknown. Most importantly these methods have not been evaluated in real-world scenarios. Some proposed LQE methods require additional hardware to work, like location devices~\cite{zhang_2009_tmc}, which are usually not available in IEEE 802.11 nodes.

Therefore, the objective is to design and evaluate, at the transport layer, a highly efficient LQE method. This method must be accurate, rapidly adaptable, environment aware, easily implementable in IEEE 802.11 devices, imposing no additional requirements on the hardware.

\subsection{Our Contribution}
\label{sec:contribution}

In this paper we propose a novel environment aware LQE method based on SNR maps and broadcast packet counts. SNR map is a profile for each data rate of IEEE 802.11b/g obtained by counting received packets for each observable SNR value. Even in the absence of traffic on a particular link the SNR map allows for accurate LQE. For broadcast packet counting we propose to use an exponentially weighted moving average (EWMA) filter to assess PDR. This significantly simplifies the implementation aspect of LQE, since only the information on the success of two recent transmissions is necessary. The two tier approach taken in our proposal, which combines the method of SNR maps and broadcast packet counts, is directed by the fact that in certain environments one approach might be better than the other, as we show later in the paper. The two tier approach is the most flexible one, however, time requirements are quite high, so we mainly use SNR maps in the rate adaptation and routing experiments in the paper. We present a simple analytical model to compare different LQE methods outlined in this paper against our proposed method. We show that our proposed SNR profile method is the most efficient and accurate one in most scenarios. A key point to note is that our method is also able to determine the weight for the EWMA filter in various environments.

Moreover, we propose a novel rate adaptation algorithm for IEEE 802.11b/g networks using our proposed LQE method. The routing layer is adapted to incorporate our accurate LQE information and fast response rate adaptation scheme to have better route selection performance. This is because the rate adaptation and route selection are two important application of the LQE. Better rate adaptation can also enhance the performance of the routing layer. Therefore, it is important to investigate the LQE with its applications together in the system level. We implement our proposed LQE and rate and route algorithms in hardware; and compare its performance using different testbeds and in diverse scenarios with the traditional beacon packet-based LQE. We show over 50\% improvement using the throughput metric based on transport layer performance assessment. To the best of our knowledge, this paper is the pioneering work which both propose better LQE and demonstrate the enhancement from the system level (including better rate adaptation and route selection) performance evaluation.

The rest of the paper is organized as follows. Section~\ref{sec:related_work} introduces the LQE classification, LQE applications, and the work related to our studies. Section~\ref{sec:comparison} presents a comparison framework of LQE methods described in Section~\ref{sec:related_work}. Section~\ref{sec:Proposed_system} describes in detail our proposed rate and route selection system with novel LQE. The testbed and the prototype used to evaluate the proposed method is introduced in Section~\ref{sec:Implementation}. The evaluation results are presented in Section~\ref{sec:Evaluation}. Finally the paper is concluded in Section~\ref{sec:Conclusion}.

\section{Classification of LQE Methods and Related Work}
\label{sec:related_work}

In Section~\ref{sec:introduction}, we have briefly explained existing link quality metrics. In this section we look at the different LQE methods closely, which are needed to compare them formally against our proposed LQE method in Section~\ref{sec:comparison}. We also review the issues related to the application of LQE in rate adaptation and routing.

\subsection{LQE Measurement}
\label{sec:LQE measurement}

LQE in IEEE 802.11 wireless networks can be categorized into four major types, based on the information sources for the estimation: packet-, SNR-, BER-based and the combination of those methods.

\subsubsection{Packet-Based LQE}
\label{sec:LQE_packet}

The direct method to assess wireless link quality is to count the number of packets received over each link in the network for a predefined unit of time. For the results of PDR properties in wireless network see~\cite{Aguayo_MCCR_2004,jardosh_2005_E-WIND-05}. Due to the different packet types used, packet-based LQE can itself be categorized into four methods: beacon packet-, broadcast packet-, unicast packet- and data packet-based.

\paragraph{Beacon Packet-Based}

The earliest method of LQE used beacon packets (40\,B broadcast packets sent at 2\,Mbps rate, also known as ``hello'' packets) to probe the link. Expected Transmission Count (ETX)~\cite{Couto_WN_2005} and Estimated Transmission Time (ETT)~\cite{BicketMobicom2005} are two beacon-based metrics. Several works reported the inefficiency of this method due to the fundamental differences between beacon packets and data packets~\cite{zhou_2009_CCNC,Kim2006,Kim_Ton_2009}. In some application, every node need to periodically send the beacon to its neighbors for other reason than LQE, if the LQE method use this beacon, we can call it passive LQE, or else we call this LQE to be active LQE method.

\paragraph{Broadcast Packet-Based}

To resolve the problem of beacon packet-based probing, nodes assessing PDR can broadcast packets equal in length to data packets, transmitted at any desired data rate\footnote{Interestingly, we were unable to find any studies that proposed and evaluated such a method in isolation.}. The overhead of this method is much higher than the beacon packet-based probing. This is because rate adaptation requires the link quality for all data rates and packet lengths. Thus many test packets must be broadcasted to cover all use cases.

\paragraph{Unicast Packet-Based}

An approach to solve the difference between broadcast probe packets and data packets transmitted in various data rates and sizes was to use unicast packets to probe the links~\cite{Kim2006}. In~\cite{fonseca_2007_hotnets} a combination of broadcast and unicast packet probing was used. When no unicast traffic was passing through the link, broadcast packets were transmitted and used for LQE. A method for very short term LQE with unicast packets was proposed in~\cite{alizai_2008_FGSN}, while in~\cite{keshavarzian_2007_communciation_letter} a mathematical model was developed to account for packet retransmissions within a single coherence time of time varying channel which could be used to predict the state of the link. We note that short term LQE is of limited use in real networks, since frequent LQE messages would saturate the routing process. In~\cite{xu_2006_Percom} authors use spatial correlation to predict the link quality and find that different PDR is observed for different locations. To implement this method in a real network, localization hardware would have to be supported by all devices. The most significant drawback of this method is the detrimental effect on network connectivity since the unicast packet will trigger more retransmissions which consume more network resources. 

\paragraph{Data Traffic-Based}

To alleviate the problems of probing with broadcast or unicast packets, authors in~\cite{zhang_2009_tmc} proposed to use the latest data packets, flowing through the link and generated by the users, for estimation of PDR. Better performance in terms of throughput is observed compared to broadcast-based probing, however, this LQE method considers only heavily loaded links. For lightly loaded links this method is ineffective and results in incomplete link state information for the network.

\subsubsection{BER-Based LQE}
\label{sec:LQE_BER}

Yet another approach to estimate the link quality is to utilize BER information~\cite{breed_2003_ber_HFE,koksal_2006_JSAC}. The estimation accuracy of this method is better than packet-based LQE since it conveys more information on the link quality and is retrieved directly from the physical layer. However, as indicated in~\cite{vlavianos_2008_PIMRC}, this method causes significant overhead since it requires processing a large amount of data. BER information is not reported by commodity wireless cards which makes this method hard to implement on standard IEEE 802.11 devices.

\subsubsection{SNR-Based LQE}
\label{sec:LQE_SNR}

This LQE method is of highest relevance in the context of this paper. The authors of ETX noted that there was a low correlation between the SNR observed over the packet counting interval and measured PDR~\cite{Couto_WN_2005}. This incorrect conclusion was a result of combining data measurements from different environments, receivers and links. It was later found that SNR can actually be a good indicator of wireless link quality~\cite{souryal_2006_wcnc,zuniga_2004_secon,srinivasan_2006_EN,lai_2003_globecom,Haratcherev_asci_2004}. The SNR-based LQE method was evaluated independently by the authors of this paper in~\cite{Zhou_VTC_2009,Zhou_Pernets_2007} and others~\cite{verma_2008_Wimesh,Zhang_Infocom_2008,Judd_Mobicom_2007,senel_2007_globecom}. However, the accuracy of these method, except for~\cite{Zhang_Infocom_2008,Judd_Mobicom_2007}, were not compared to any other estimation method. Moreover, except for~\cite{senel_2007_globecom}, a static SNR to PDR mapping profile was used, thus adaptation to changing environments was not considered. Authors in~\cite{wang_2006_realman} proposed different metrics that correlated with packet delivery probability and used machine learning techniques to select the best metric in the LQE process. Nevertheless, the fact that different rates in IEEE 802.11 have different SNR profiles, as pointed out in~\cite{Zhou_VTC_2009,Zhang_Infocom_2008}, was not taken into account in the analysis.

A separate discussion is needed for~\cite{Zhang_Infocom_2008}, which is closest to our work. Although a correlation between SNR and PDR was well discussed, a piecewise linear approximation of this relation was used in rate adaptation process, see~\cite[Fig. 7]{Zhang_Infocom_2008}, which loses throughput gain in transitional regions of SNR map~\cite[Fig. 3]{zuniga_2004_secon}. Also, to obtain the SNR map, the authors required a calibration process in which, unrealistically, an interference-free channel was assumed. Further, the profiles were obtained by probing the link with very high intensity, i.e. 20\,ms of each second, which creates huge overhead for the device. But the most significant drawback of this method is the usage of beacon packet of different size than data packet to obtain SNR profiles, which as we pointed out earlier, results in inaccurate rate adaptation.

\subsubsection{Combined Methods of SNR- and Packet-Based LQE}
\label{sec:LQE_combined}

Since, individually, SNR- and packet-based LQE have advantages and disadvantages, authors in~\cite{vlavianos_2008_PIMRC} remarked to combine them for use depending on the environment and LQE accuracy. Works in which such combined methods were proposed and evaluated independently were~\cite{zhou_2009_CCNC,Fonseca_Hotnets_2007,Zhou_Pernets_2007}. The estimator was a weighted function of the individual packet- and SNR-based LQEs. However, the metric had limited practical application since the weight had to be set individually for each environment and data rate, among others, and the learning process for the optimal weight involved the entire history of network operation.

A qualitative comparison of all existing LQE methods is given in Table~\ref{LQE_comparision}. As noted earlier, in Section~\ref{sec:comparison} we propose a simple analytical framework to compare them quantitatively.

\begin{table}
\centering
\caption{Qualitative Comparison of LQE Methods (Desired Performance: Update Ability--High, Accuracy--High, Overhead--Low)}
\label{LQE_comparision}
\begin{tabular}{c|c|c|c}
  \hline
  Information Source & Update Ability & Accuracy & Overhead \\
  \hline\hline
  Packet (beacon)~\cite{Couto_WN_2005,BicketMobicom2005} & Low & Low & Low \\
  \hline
  Packet (broadcast) & Low & Medium & High \\
  \hline
  Packet (unicast)~\cite{Kim2006,Kim_Ton_2009,alizai_2008_FGSN,xu_2006_Percom} & Low & Medium & High \\
  \hline
  Packet (data)~\cite{zhang_2009_tmc} & High & High & High \\
  \hline
  SNR~\cite{verma_2008_Wimesh,senel_2007_globecom,Judd_Mobicom_2007, Zhou_Pernets_2007, Zhou_VTC_2009,Zhang_Infocom_2008} & High & Medium & Low \\
  \hline
  BER~\cite{breed_2003_ber_HFE,koksal_2006_JSAC} & Medium & High & High \\
  \hline
  Packet and SNR~\cite{zhou_2009_CCNC} & Low & Medium & Low \\
  \hline
\end{tabular}
\end{table}

\subsection{Link Quality-based Applications}
\label{sec:lqe_applications}

LQE is required for rate adaptation and dynamic or ad hoc routing in wireless mesh and wireless sensor networks. We consider each separately in relation to the link quality.

\subsubsection{Rate Adaptation}

Earliest works like~\cite{Holland_Mobicom_2001,Lacage_MSWIM_2004} proposed to use a basic packet counting method to decide the data rate, which, as we have explained in Section~\ref{sec:LQE_packet}, is very problematic. Beacon packet-based counting was used in~\cite{Bicket_SampleRate_2005} to determine the optimal data rate. This resulted in stale and inaccurate link quality information for all types of packets. Authors in~\cite{Wong_Mobilcom_2006} claimed that they achieved better performance than~\cite{Bicket_SampleRate_2005}, though the problem of packet based LQE has not been solved. The SNR profile was used for rate adaptation first in~\cite{Judd_Mobicom_2007,Zhang_Infocom_2008}. However, in~\cite{Judd_Mobicom_2007} SNR was estimated per packet level which is not suitable for routing algorithms, while in~\cite{Zhang_Infocom_2008} due to the inefficiency of the SNR profile creation, as pointed out in Section~\ref{sec:LQE_SNR}, the proposed rate adaptation can be easily outperformed. Also, rate adaptation~\cite{Zhang_Infocom_2008} was evaluated in indoor environment only and no mobility in rate adaptation process was taken into account. It is interesting to note that even with rather inefficient rate adaptation process it performed better than other well known rate adaptation algorithms~\cite{Holland_Mobicom_2001,Sadeghi_mobicom_2002,Chen_infocom_2007}.

\subsubsection{Route Selection}

As in rate adaptation, most of the link quality-based routing methods are based on link probing with broadcast packets~\cite{Couto_WN_2005,BicketMobicom2005}. In these works authors show that the routing performance is better than the hop-count based route selection, where the degree of improvement differed depending on the propagation environment. Authors in~\cite{koksal_2006_JSAC} proposed two metrics called modified ETX (mETX) and Effective Number of Transmissions (ENT), which represent the variability of steady state packet error probability and effective bandwidth (in the network sense), respectively. The metrics are designed to enable lower layers reduce the loss rates visible to the upper layer protocols like TCP. A single data rate is considered in~\cite{koksal_2006_JSAC} and the metrics are evaluated in simulation environments only. Authors in~\cite{tsai_2006_wpc} modified the traditional AODV routing protocol and used SNR to select routes directly. The proposed method decides whether particular links should be eliminated at the routing layer based on the selected SNR threshold. Again, the protocol was evaluated only via computer simulation. In~\cite{karbaschi_2008_wowmom} broadcast packet-based traffic was used to evaluate link quality and select a route. Apparently, the problems of packet counting-based methods remained. Finally, the method presented in~\cite{wang_2006_realman} does not give a comparison to other known route selection methods and the routing itself is performed in a single data rate. We note that the rate adaptation method based on SNR maps of~\cite{Zhang_Infocom_2008} was not evaluated on the routing layer. In addition to the aforementioned issues, none of the above works discussed a mechanism to update the estimation parameters of used LQE method, which is of utmost importance for large scale mesh networks.

\section{A Simple Analytical Assessment of the Proposed LQE}
\label{sec:comparison}

Our novel LQE system is based on a combination of two methods, as noted in Section~\ref{sec:contribution}, i.e., SNR maps and broadcast packets using the same data rate and packet size as application packets. While broadcast packet-based LQE generally conforms to the well described process of LQE based on counting received packets, the novelty of our LQE method lies in the way of selecting these methods, their update, and generating SNR maps in a different approach. While the detailed discussion on overall LQE system, proposed rate and route adaptations using our LQE method and the overall system implementation in networking hardware is given in Section~\ref{sec:Proposed_system}, we  briefly describe here the idea behind creating SNR maps and their usage in LQE. This is needed to evaluate it analytically and compare against considered LQE methods described in Section~\ref{sec:related_work}. Analytical model introduced here will allow to show using simple expressions the source of impairments in LQE of each described above method.

\subsection{Proposed SNR map-based LQE: General Overview}
\label{sec:snr_maps_general_idea}

In our implementation of SNR maps-based LQE, each node counts the number of acknowledged data packets from intended receivers for each SNR value and data rate it reads from the packet. Based on these counts, each node creates a map, i.e. a relation between all observed SNR values and the associated number of successfully received packets. In our method the map is updated whenever new data traffic passes between considered transmitted and receiver. Please note that we do not transmit any probing packets, perform any calibrations initially or linearize SNR maps, as other methods do, see Section~\ref{sec:LQE_SNR}. We constantly update the maps whenever traffic passes and use the actual data packets (not probing packets) to create maps. This way there is no impairment between SNR maps and the actual SNR to PDR relation and the mapping is always up-to-date. The impairment is usually a result of linearizing the mapping based on some mathematical transformations.

We will now compare our proposed SNR map-based LQE method with other methods listed in the previous section. Specifically, we will compare unicast, beacon, broadcast, and data packet-based LQE, against proposed LQE based on SNR maps, referred with the variable subscripts $u$, $b$, $r$, $d$ and $s$, respectively. We will not compare combined LQE methods, since we focus on each method individually, or methods based on BER, due to their impracticality, as remarked in Section~\ref{sec:LQE_combined}.

\subsection{System Model}
\label{sec:system_model}

We assume that each node in the network is located at a fixed position. The statistical process behind signal strength (given in linear scale) $\psi\sim \mathcal{N}(\Psi,\sigma)$ is stationary and mimicks an AWGN channel. Each node has $\bar{N}=g_t(N)$ number of neighbors, where $N$ is the number of nodes in the network and $g_t(\cdot)$ is the network topology~\cite[Ch. 3]{Hekmat_phd_thesis}. Any rate $R=\{1, 2, 5.5, 6, 9, 11, 12, 18, 24, 36, 48, 54\}$\,Mbps of IEEE 802.11b/g is allowed for the nodes to transmit. Nodes send data packets of $A_d$ Bytes transmitted at any rate $R$ and beacon packets of $A_b$\,Bytes are sent at rate $R(2)$. Note that based on compatibility requirement, for the higher data rate standard, e.g. IEEE 802.11g, the device will still use the low data rate packets to associate with other nodes, since in most of current work on LQE $R(2)$ is assumed as the beacon rate, which is used to associate with neighbor nodes. During each link estimation interval for all LQE methods, except for data packet-based and SNR profile-based LQE, $w$ packets are sent in equal intervals. On the contrary data packets are sent randomly at an average rate of $w_d$ per estimation interval per node neighbor.

We will use two metrics to compare the LQE methods: (i) time it takes to estimate the link quality and (ii) efficiency defined as a distance from actual relation between SNR and PDR. We will not take update ability into consideration, which is qualitatively assessed in Table~\ref{LQE_comparision}. See Table~\ref{variables} for the list of variables used in this section and in the remaining parts of the paper.

\subsection{Estimation Time Analysis}

The time spent to transmit packets in one estimation interval for unicast packet-based LQE is~\cite[Sec. II-A]{verma_2008_Wimesh}:
\begin{equation}
\tau_u=\left(\sum_{i=1}^{|R|}\frac{A_{d}}{R(i)}+\left[t_{di}+t_{si}+t_{bt}+\frac{A_b}{R(2)}\right]|R|\right)w\bar{N},
\end{equation}
where $t_{di}$ is the length of IEEE 802.11b/g Distributed Inter-Frame Space, $t_{si}$ is the length of IEEE 802.11b/g Short Inter-Frame Space, $t_{bt}=\frac{c_{m}}{2}t_l$ is the average backoff time\footnote{In the implementation of SNR maps-based LQE, described in Section~\ref{sec:Proposed_system}, we will take PDR into account while evaluating $t_{bt}$.} with a uniformly distributed contention window size with minimum value of $c_{m}$ and slot size $t_l$, and $|R|=12$ is the size of $R$.

For beacon packet-based LQE average time consumed in one estimation interval is
\begin{equation}
\tau_b=\left(\frac{A_b}{R(2)}+t_{si}+t_{bt}\right)w\bar{N},
\end{equation}
and in the case of broadcast packet-based LQE consumed time for the estimation is
\begin{equation}
\tau_r=\left(\sum_{i=1}^{|R|}\frac{A_d}{R(i)}+[t_{si}+t_{bt}]|R|\right)w\bar{N}.
\end{equation}
Finally for the data packet-based LQE and SNR profile-based LQE consumed time is
\begin{equation}
\tau_d=\tau_s=\left(\bar{T_d}+t_{di}+t_{si}+t_{bt}+\frac{A_b}{R(2)}\right)w_d\bar{N_p},
\end{equation}
where $\bar{T_d}=\frac{1}{|R|}\sum_{i=1}^{|R|}\frac{A_{d}}{R(i)}$ is the mean data packet transmission time across all data rates, and $\bar{N_p}=\bar{N}p_a$ is the mean number of neighbors to whom each node transmits, with the probability $p_a$ to any of $\bar{N}$ neighbors.

We finally define the inverse time consumption metric normalized to $\tau_d$ for each LQE method
\begin{equation}
\epsilon_{t,x}=\frac{\tau_d}{\tau_c}\in[0,1],
\end{equation}
where
\begin{equation}
\tau_c=
\begin{cases}
\tau_d, & \text{data packet and SNR profile},\\
\tau_{x=\{u, b, r, s\}}+\tau_d, & \text{otherwise}.
\end{cases}
\end{equation}

\subsection{LQE Efficiency Analysis}

There are three essential factors that influence LQE efficiency: (i) SNR estimation accuracy, (ii) validity of the mapping between SNR to PDR, and (iii) the difference between the probing and data packet. We can therefore construct the PDR estimation efficiency metric, $\delta_{x}$ for each LQE method as
\begin{equation}
\delta_{x}=\sum_{k=1}^{|R|}\left|\underset{M_{x}}{\underbrace{\left(1-f_{e,k}\left(\Psi_x\right)\right)^{A_e}}}-\underset{M_y}{\underbrace{\left(1-f_k\left(\Psi\right)\right)^{A_{d}}}}\right|,
\label{eq:lqe_diff}
\end{equation}
where  $x=\{d, u, b, r, s\}$, $f_k(\cdot)$ is the actual mapping function between SNR and PDR for a data packet, $f_{e,k}(\cdot)$ is the mapping used by LQE method $x$, $\Psi_x$ is the mean estimated SNR by LQE method $x$, and $A_e$ is the size of the probing packed used by the LQE process $x$. Note that (\ref{eq:lqe_diff}) is a metric that compares the average (summed over all rates in $R$) estimation efficiency as a difference between the actual SNR and PDR relation for a given rate and packet size and a relation obtained using an estimation method of interest.

For unicast and broadcast packet-based LQE
\begin{equation}
M_{x=u}=M_{x=r}=\left(1-f_k(\Psi_c)\right)^{A_d}
\label{eq:mu}
\end{equation}
and for beacon packet-based LQE
\begin{equation}
M_{x=b}=\left(1-f_{2}(\Psi_c)\right)^{A_b}.
\label{eq:mb}
\end{equation}
Please note that in (\ref{eq:mu}) and (\ref{eq:mb}) the same estimated SNR $\Psi_c=\Psi_{\{u,r,b\}}$ is observed, since each method uses the same number of packets from which the SNR $\psi$ is extracted and averaged. In case of data packet-based LQE
\begin{equation}
M_{x=d}=\left(1-f_k(\Psi_d)\right)^{A_b}.
\label{eq:md}
\end{equation}
Before we compute $M_{x=s}$ for the SNR profile-based LQE we introduce the following proposition.
\begin{proposition}
For SNR profile-based LQE $M_{x=s}=M_y$ asymptotically.
\begin{proof}
For all but the SNR profile-based LQE method minimum squared error of the SNR estimate is,
\begin{equation}
\Psi_{e,x}=
\begin{cases}
\mathbb{E}[(\Psi-\Psi_d)^2]=\frac{\sigma}{\bar{w}R}, & \text{data packet-based LQE},\\
\mathbb{E}[(\Psi-\Psi_c)^2]=\frac{\sigma}{w}, & \text{otherwise}.\\
\end{cases}
\end{equation}
Again, note that $\Psi_{e,x}$ is always constant for every estimation interval, since with every interval each LQE method, except for SNR profile-based, will discard previous measurements and estimate current SNR based on recently received packets. However, in the case of the SNR profile the measurements are never discarded. The profile is always updated after every new measurement interval. The estimation error in this case is
\begin{equation}
\Psi_{e,x}=\mathbb{E}[(\Psi-\Psi_s)^2]=\lim_{\bar{w}\rightarrow\infty}\frac{\sigma}{\bar{w}R}=0
\end{equation}
which results in estimated SNR value for SNR profile-based LQE $\Psi_s=\Psi$. Since $M_s$ is constructed in the same way as (\ref{eq:md}) but replaced with accurate SNR it completes the proof. \end{proof} \label{lem:1} \end{proposition}

Finally, the normalized estimation efficiency metric for method $x$ is
\begin{equation}
\epsilon_{e,x}=1-\frac{\delta_{x}}{\sum_{x=\{d, u, b, r, s\}}\delta_x}\in[0,1].
\end{equation}

\subsection{Overall LQE Efficiency}

Having $\epsilon_{t,x}$ and $\epsilon_{e,x}$ we define an overall efficiency metric as $\epsilon_{x}=\epsilon_{e,x}+\epsilon_{t,x}\in[0,2]$.

This metric allowed us to assess each method and obtain the information on which feature (time or efficiency) was more dominant in the estimation process. In the next section we present an example numerical results of the comparison.

\begin{table*}
\centering
\caption{Summary of Used Variables}
\label{variables}
\begin{tabular}{c|l|c}
  \hline
  Value & Description & Unit\\
  \hline\hline
  $c_m$ & minimum window size & $\mu$s \\
  $g_t(\cdot)$ & neighbor function & --- \\
  $f_{k,e}(\cdot)$, $f_k(\cdot)$ & estimated and actual BER function for rate $k$ & --- \\
  $k$, $i$ & rate index, time instant & --- \\
  $p_a$ & probability of connecting to neighbor & --- \\
  $r_k$ & index of the selected rate & --- \\
  $t_l$ & slot length & $\mu$s\\
  $t_{si}$, $t_{di}$ & distributed and long interframe space & $\mu$s \\
  $t_{p,R(k)}$ & packet transmission time in lossless condition & $\mu$s \\
  $t_{b,R(k),i,\Psi}$ & backoff time given retransmissions & $\mu$s \\
  $w$, $w_d$ & number of probes and data packets & --- \\
  $w(t)$ & number of acknowledged packets in time $t$ & --- \\
  $A_d$, $A_b$ & size of data and beacon packets & kB \\
  $D_m$ & MAC address & --- \\
  $K$ & Rician factor & dB \\
  $M_x$, $M_y$ & probe and actual SNR to PDR function & --- \\
  $N$, $\bar N$, $\bar N_p$ & number of nodes, neighbors and connected neighbors & --- \\
  $P$, $S$ & size of one and complete SNR profile & Bytes \\
  $\mathcal{R}$ & routing metric & --- \\
  $R(x)$ & IEEE 802.11b/g rate $x$ & Mbps\\
  $\bar T_d$ & mean transmission time of $A_d$ & $\mu$s \\
  $E_{R(k),(i)}$, $E_{R(k),(i),d}$, $E_{R(k),(i),\Psi}$ & estimated PDR using broadcast packets, data traffic and SNR map& --- \\
  $P_{R(k),i}$ & PDR in SampleRate & --- \\
  $F_{(i-1,i),R(k)}$ & number of retransmissions & --- \\
  $G_{R(k),i,\Psi}$ & rate metric & --- \\
  $X_{R(k),(i),\Psi}$ & packet reception indicators & --- \\
  $\mathbf{E}_{\Psi}$ & vector of PDRs & \\
  $M_{\Psi,k,D_m}$ & data structure with $tx$ and $t_{ack}$ for $\Psi$, $k$, $D_m$ & --- \\
  $\alpha_p$, $\alpha_s$, $\alpha_r$ & packet, SNR and rate smoothing factors & --- \\
  $\sigma$ & SNR variance & dBm \\
  $\epsilon_{t,x}$, $\epsilon_{e,x}$, $\epsilon_{x}$ & normalized time, estimation and total efficiency & --- \\
  $\tau_x$, $\delta_x$ & time and estimation efficiency of metric $x$ & --- \\
  $\Delta$ & PDR estimation accuracy & --- \\
  $\psi$, $\Psi$, $\Psi_{x=\{d, s, c\}}$, $\Psi_{e,x}$ & SNR: instantaneous, mean, individual ($d$ data packet, $s$ SNR profile, $c$ rest), error & dBm \\
  $\mathbf{\Psi}$ & vector of SNRs & dBm\\
  $\Phi(\cdot)$ & Normal distribution CDF & --- \\
  \hline
\end{tabular}
\end{table*}

\subsection{Numerical Evaluation}

\begin{figure}
\centering
\includegraphics[width=0.32\columnwidth]{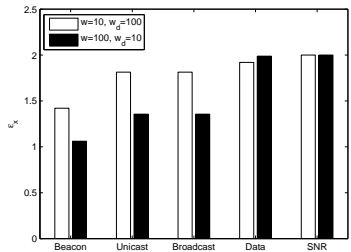}
\caption{Numerical comparison of available LQE methods for two different sets of $w$ and $w_d$ in AWGN channel. Assumed parameters: $A_d=1500$\,kB, $A_b=40$\,B, $N=40$, $t_{si}=10$\,$\mu$s, $c_m=32$, $t_l=20$\,$\mu$s, $p_a=0.3$, $\Psi=10$\,dBm, $\delta=1$\,dBm, $R=\{1,2\}$\,Mbps.}
\label{fig:method_comp}
\end{figure}

As an example we will focus on a tandem network in which $\bar{N}=\frac{2(N-1)}{N}$~\cite[Ch. 3]{Hekmat_phd_thesis} (note that the analysis can be easily extended to any network topology). Results are given in Fig.~\ref{fig:method_comp} with the assumed parameters therein. We remark that, without loss of generality, we have focused only on $R=\{1, 2\}$\,Mbps for which $f_{k}(\cdot)$ in AWGN channel, used in this evaluation, are given as $f_{1}(\Psi)=Q(\Psi)$, for BPSK, and $f_{2}(\Psi)=Q(\Psi/2)$, for QPSK, respectively, where $Q(\cdot)=1-\Phi(\cdot)$ and $\Phi(\cdot)$ is a CDF of a Normal distribution. The reason for focusing on two rates only was the simple and tractable form of the $f_{k}(\cdot)$ function for these two rates. In the numerical evaluation we have assumed that $\Psi_x=\Psi-\Psi_{e,x}$. Note that both sets of $w$ and $w_d$ are chosen such that one sending rate is greater than another.

We see that the SNR profile-based LQE performs best irrespective of $w$ and $w_d$. Beacon packet-based LQE has the worst performance. Note that because of the increased accuracy of probing with beacons ($w=100$), overall efficiency is still low due to high overhead of the method. Also note that unicast and broadcast packet-based LQE perform equally well, irrespective of $w$ and $w_d$ value. The difference in $\epsilon_x$ between data packet-based and SNR profile based LQE is greater when $w_{d}\gg w$.

\section{Proposed Link Quality Aware IEEE 802.11 Network: Detailed Description}
\label{sec:Proposed_system}

Knowing in general what affects known LQE processes and showing that our proposed method outperforms other existing methods, we now introduce in detail proposed LQE process from the perspective of link quality aware IEEE 802.11b/g ad hoc network. We will describe in detail procedures used in evaluating link quality in the considered class of networks and the novel method for rate adaptation and routing which utilizes the proposed LQE method. Later in Section~\ref{sec:Evaluation}, we will present the results of the evaluation of our LQE method, given in Section~\ref{sec:Implementation}.

\subsection{Proposed LQE Method}
\label{sec:lqe_proposed}

\begin{figure}
\centering
\subfigure[]{\includegraphics[width=0.49\columnwidth]{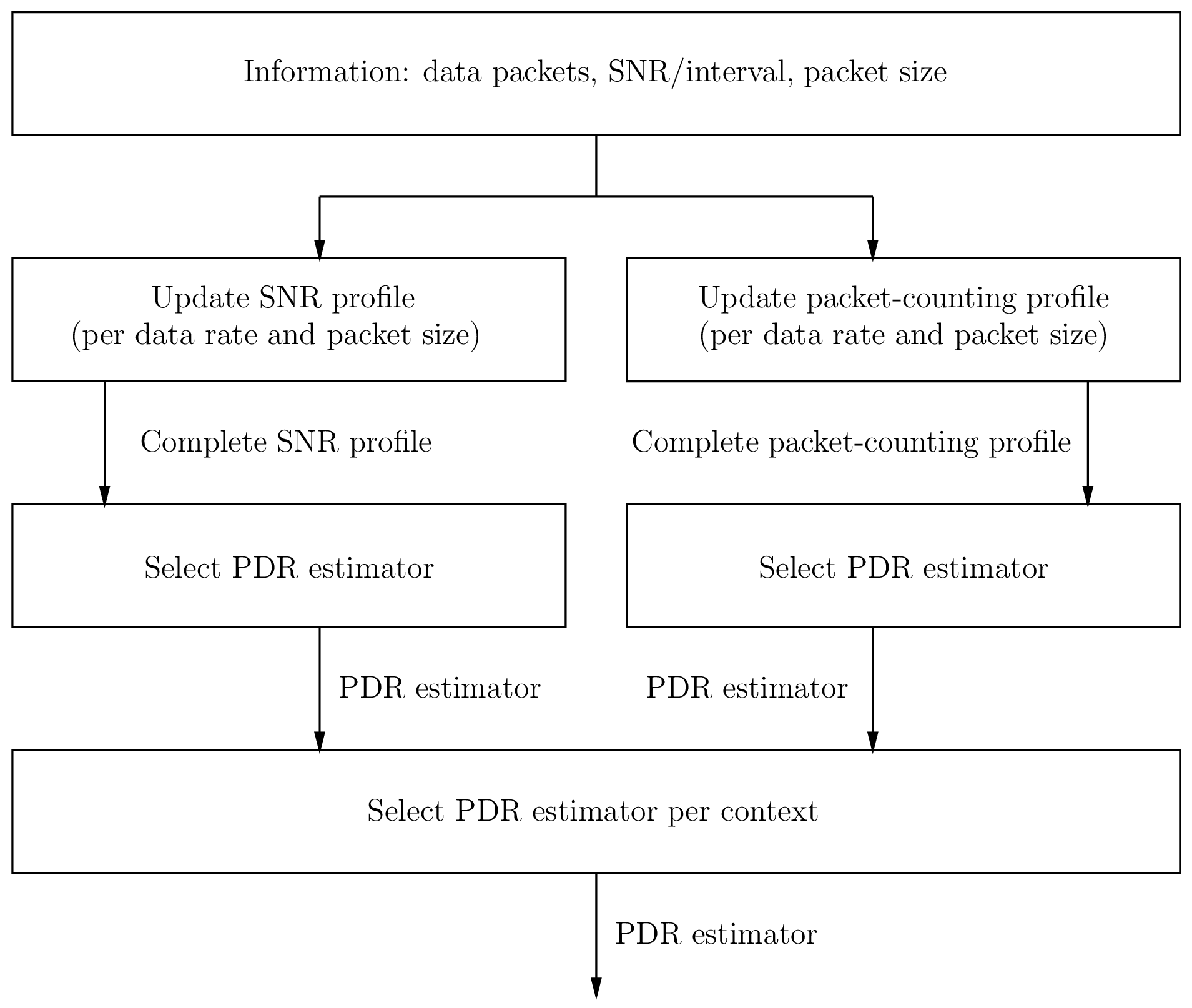}\label{Fig_combined}}
\subfigure[]{\includegraphics[width=0.49\columnwidth]{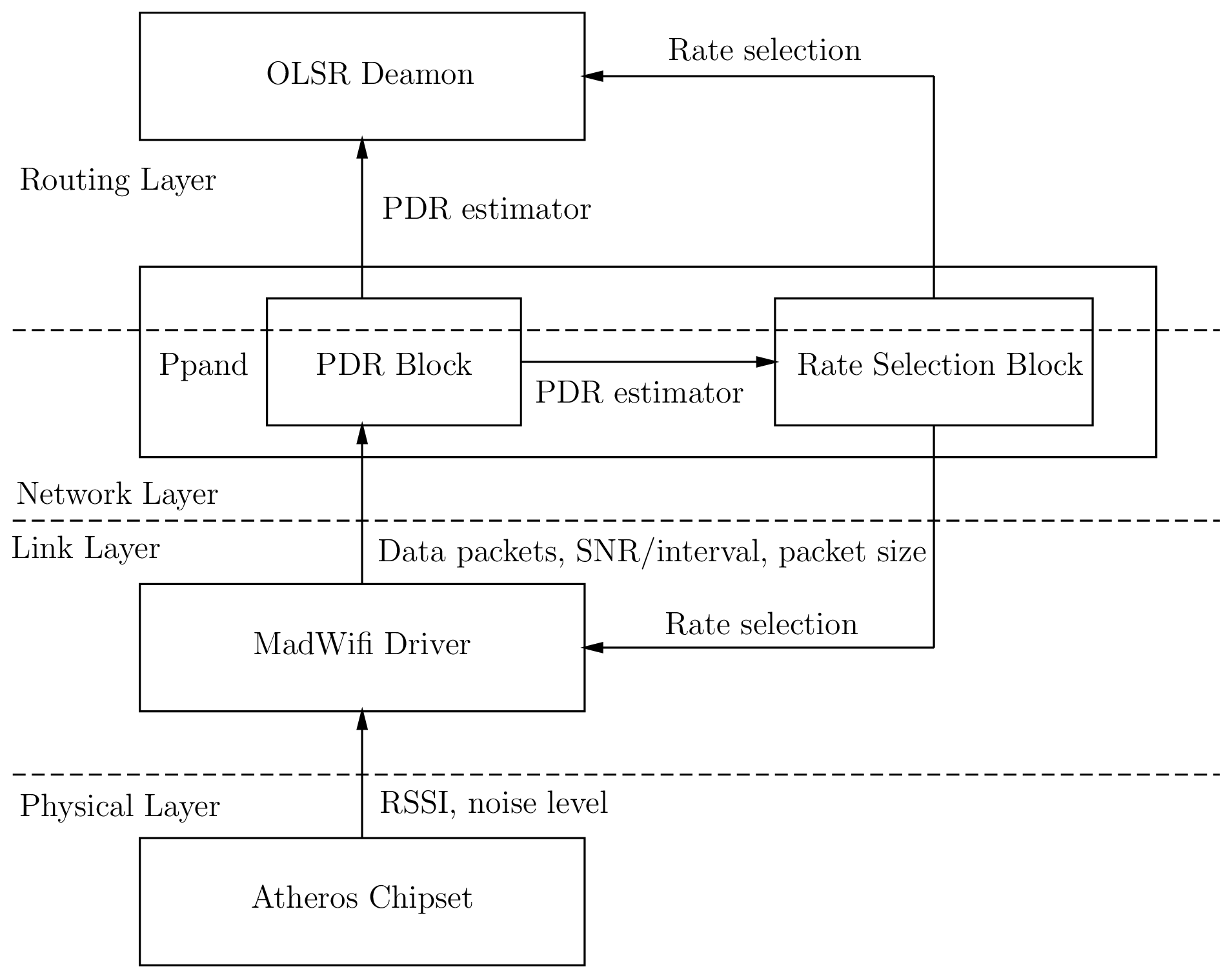}\label{Fig_implementation}}
\caption{Proposed IEEE 802.11b/g system: (a) LQE system (PDR block), and (b) software implementation of the complete IEEE 802.11b/g network node, see also~\cite[Fig. 1]{zhou_2009_CCNC} for a similar implementation.}
\end{figure}

In our approach we measure link quality with PDR. The proposed LQE is based on two independent methods of evaluating PDR: (i) SNR profile and (ii) received broadcast packet counts. Depending on the environment and required LQE efficiency the appropriate method will be chosen. The proposed LQE flow graph is depicted in Fig.~\ref{Fig_combined}.  We will describe and discuss each method in detail.

\subsubsection{Packet Counting}
\label{sec:packet_counting}

Every second we transmit a fixed size broadcast packet of 1.5\,kB for each supported data rate. For each data rate every receiver calculates its PDR using a EWMA filter
\begin{equation}
E_{R(k),i}=\alpha_p X_{R(k),i}+(1-\alpha_p)E_{R(k),i-1},
\label{eq:eirkp}
\end{equation}
where $E_{R(k),i}$ is the estimated PDR for rate $k$ at discrete time instant $i$, $X_{R(k),i}=1$ flags a correctly received broadcast packet at time $i$ and rate $R(k)$, while $X_{R(k),i}=0$ indicates otherwise, and $\alpha_p$ is the smoothing factor set for each network environment.

It is important to note the differences between our packet counting LQE approach and the LQE used in calculating ETX~\cite[Eq. (1)]{Couto_WN_2005} and ETT~\cite{BicketMobicom2005} metrics. First, our method requires only one packet sent per time interval to obtain a PDR estimate, while LQE in ETX requires $w$ broadcasting intervals (in~\cite{Couto_WN_2005,BicketMobicom2005} $w=10$) to calculate delivery ratio. Second, our method continuously updates the PDR, while LQE of ETX discards previous measurements every $w$ broadcasts. To emphasize, we are using the packet with the same size as data packet while traditional LQE in ETX, while ETT use beacon packet. We replace the LQE method in the ETX or ETT method and therefore achieved even better performance. In our LQE, the estimation is per each neighbor, which means each link is estimated by the nodes at each side of the link. If the link is reciprocal, there will be two estimation result. The LQE results are used in the rate adaptation process for each node. Therefore, the reciprocal links will have no effect on the result for the LQE, as well as for the rate adaptation part and routing. 

\subsubsection{SNR Profile}
\label{sec:snr_profile}

After an initial node bootstrap, a node sends unicast packets of 1.5\,kB at 10 packets per second, and at every IEEE 802.11b/g data rate in $R$ over each link. During the period $t$ each node is able to send $w(t)$ packet transmission attempts, while the sender waits for every acknowledgment. For one hour of link probing we calculate PDR as
\begin{equation}
E_{R(k),\Psi}=\frac{1}{w(t)}\sum_{i=1}^{w(t)}X_{R(k),i,\Psi},
\label{eq:e_sj}
\end{equation}
where $E_{R(k),\Psi}$ is the mean PDR for discrete SNR value $\Psi$ and data rate $R(k)$, $X_{R(k),i,\Psi}=1$ when a unicast packet was received at discrete time instant $i$, SNR $\Psi$ and rate $R(k)$, and $X_{R(k),i,\Psi}=0$, otherwise. This way we create a mapping relation between SNR and PDR, which we refer to as SNR map or SNR profile, for all IEEE 802.11b/g rates. This procedure is needed for devices that have no SNR maps, such that device deployed elsewhere will be able to easily update its existing map without necessary probing. If some devices are power limited, the bootstrap period can be omitted. In this case the device can first use an universal SNR map saved in the system to start up. This way the map can still be updated by utilizing or overhearing the passing by traffic. SNR values $\Psi$ are registered once per second, i.e., once per 10 transmitted unicast packets. Ideally a profile should be created for each packet size. It is important to note however, that in the current Internet approximately 50\% of the traffic is sent with the Maximum Transmit Unit (MTU) size, where the MTU for IPv6 is 1.268\,kB, for IEEE 802.3 1.492\,kB, for Ethernet II 1.5\,kB, and for IEEE 802.11 2.272\,kB~\cite{Williamson_ic_2001}\footnote{It was also noted in~\cite{Williamson_ic_2001} that about 40\% of packets are 40\,B long, which accounts for TCP ACK packets. The remaining 10\% of packets are scattered between two extremes, i.e ACK and MTU.}. In our implementation we chose 1.5\,kB packets, since the majority of Internet traffic is of wired network provenience and wireless network gateways relay packets unaltered from the wired domain.

Due to constant changes in the propagation conditions, as well as possible changes in the network environment, results of the initial measurements must be updated. If and only if a node sends data traffic with a given packet size is the measurement result updated as
\begin{equation}
E_{R(k),i,\Psi}=\frac{\alpha_s}{w(t)+1}\sum_{l=1}^{w(t)+1}X_{R(k),l,j}+(1-\alpha_s)E_{R(k),i-1,\Psi},
\label{eq:e_sji}
\end{equation}
where $E_{R(k),i,\Psi}$ is the estimated SNR profile mapping point for SNR $\Psi$ at discrete time instant $i$ and rate $R(k)$ and $\alpha_s$ is the smoothing factor, which is specific to the network environment. Equation (\ref{eq:e_sji}) smoothens the incoming measurements of $E_{R(k),i,\Psi}$ with a EWMA filter, just as in~(\ref{eq:eirkp}). The complete SNR profile for one node and one data rate is then a $n$-tuple
\begin{equation}
E_{R(k),\Psi}=\{E_{R(k),i,\mathbf\Psi(1)},\ldots,E_{R(k),i,\mathbf\Psi(n)}\},
\label{eq:eirks}
\end{equation}
where $\mathbf\Psi=\{\Psi_1,\ldots,\Psi_n\}$ is a $n$-tuple of all SNR values reported by the physical layer. Note that due to the link reciprocity assumption the same profile is used for reverse and forward direction of a link. The profile occupies $S=P|R|\bar{N}$\,B of uncompressed disk space, where $P$ is the size in Bytes of individual SNR profile for each packet size (in our case $P=8$\,kB). For example, in the case of $\bar N=10$, $S=960$\,kB.

\subsubsection{Method Selection Based on Network Environment and Required LQE Accuracy}
\label{sec:context}

Environment (or more generally context) awareness can indeed improve the LQE process. Context encompasses operating state of a device: whether it is indoor or outdoor, moving or not, whether the communication has line of sight (LOS) or not, and what level of LQE accuracy is required by the applications. Information about the environment can be input by the user or detected by the device itself. In the latter case the node must be location aware, which may require an additional hardware. In the proposed LQE system, see Fig.~\ref{Fig_combined}, context is needed to decide which PDR method to use at a given moment. We show in Section~\ref{sec:lqe_evaluation} that in very limited scenarios packet based counting achieves better performance than SNR profile-based LQE. Context, however, is not absolutely necessary, since the SNR profile can be updated, which leads to a comparably accurate LQE after some learning time. This will also be shown in Section~\ref{sec:lqe_evaluation}. Our conclusion about this selection is that if the system require very high LQE accuracy, the selection procedure will bring in the highest accuracy since this is the combination of the both method. However, in most cases, the SNR mapping will provide enough accuracy for most of applications, such as rate adaptation and routing.

\subsection{Improved Rate Selection Mechanism}
\label{sec:SNR_RATE}

We propose to augment the classic SampleRate rate adaptation protocol~\cite{Bicket_SampleRate_2005} with our LQE method. SampleRate was a convenient choice, since it was implemented in the IEEE 802.11 open source driver Madwifi~\cite{MadWifi} which was used in our algorithm implementation, see Section~\ref{sec:Implementation}.

The SampleRate protocol works as follows. For a given packet size every node picks a link with a data rate such that
\begin{equation}
\underset{R(k)}{\arg \min}\left\{\frac{A_d}{P_{R(1),i}R(1)},\ldots,\frac{A_d}{P_{R(12),i}R(12)}\right\},
\label{eq:minrate}
\end{equation}
where $P_{R(k),i}$ is the PDR for data rate $R(k)$ at discrete time instant $i$ defined as
\begin{equation}
P_{R(k),i}=\frac{\alpha_r}{1+F_{(i-1,i),R(k)}}+(1-\alpha_r)P_{R(k),i-1}.
\label{eq:prki}
\end{equation}
In~(\ref{eq:prki}), $F_{(i-1,i),R(k)}$ is the number of retransmissions that occurred for one packet during the $(i-1,i)$ interval and at data rate $R(k)$ and $\alpha_r$ is the smoothing factor optimized for a given environment. In SampleRate, 10\% of packets sent per link are transmitted at a data rate other than the selected rate (see~\cite[Ch. 5]{Bicket_SampleRate_2005}) to increase the channel's probing performance and improve rate selection. The rate selection process using the SampleRate algorithm is effective only when enough traffic passes through the link since in~(\ref{eq:prki}), $F_{(i-1,i),R(k)}$ is computed only with the acknowledgments of the transmitted packets.

In the proposed rate selection process during each time instant $i$, given SNR $j$, we compute the following rate metric
\begin{equation}
G_{R(k),i,j}=\frac{1}{E_{R(k),i,j}}t_{p,R(k)}+t_{b,R(k),i,j},
\end{equation}
where $t_{p,R(k)}=\frac{A_d}{R(k)}$ is the time required to transmit a packet of $A_d$ under lossless conditions and
\begin{equation}
t_{b,R(k),i,j}=\frac{c_m}{2}\left(1+\sum_{i=0}^{10}2^i[1-E_{R(k),i,j}]\right)
\end{equation}
is the backoff time for rate $R(k)$, SNR $j$, and time instant $i$ which takes link quality into consideration~\cite[Eq. 17]{Draves_MobiCom_2004}. We assume a maximum 11 retransmissions. Finally our proposed algorithm chooses the rate such that $\underset{R(k)}{\arg \min}\left\{G_{R(1),i},\ldots, G_{R(12),i}\right\}$.

The complete implementation of our rate adaptation process is given in Algorithm~\ref{alg:1}. Please note that we omitted index $i$ for clarity and we present the algorithm for SNR map only when the initial SNR profile for all data rates is available, i.e., we omit the initial phase to compute~(\ref{eq:e_sji}).

It is important to note that, like the original SampleRate, we deviate randomly from the chosen data rate to increase link probing efficiency. For each interval, 0.5\% of packets are sent at data rate $R(k-1)$ and 0.5\% at data rate $R(k+1)$, where the chosen rate is $R(k)$. In case $k=1$ only $R(2)$ is used while in case $k=12$ only $R(11)$ is used. The percentage of packets sent at adjacent rates was selected empirically such that protocol performance was maximized. Any increase of the probing rate resulted in significant decrease in network throughput. Also, we decided to probe only adjacent data rates, since probing more data rates caused instabilities in the routing process.

\subsection{Improved Route Selection Mechanism}
\label{sec:SNR_ROUTE}

The proposed link quality metric can be applied in the network layer to the routing protocols. By using a better LQE technique in the route selection process, it is possible to gain throughput over the links. We consider a link-state routing protocol in the implementation process described in Section~\ref{sec:Implementation}. In link-state routing protocols, the ETX per link is calculated using~\cite[Eq. 1]{Couto_WN_2005} for which the PDRs per neighbor link are collected through topology control (TC) messages. We propose the following routing metric per link in the TC
\begin{equation}
\mathcal R=\frac{1}{E_{R(j),i}\sqrt{R(j)}}\frac{1}{E_{R(k),i}\sqrt{R(k)}},
\label{eq:M}
\end{equation}
which replaces ETX and is very similar to ETT.

Using $\mathcal R$, the link with the highest throughput and PDR is always selected. To balance the accuracy of LQE and the number of route update messages we execute one LQE interval per routing update interval.

\begin{algorithm}
%\linesnumbered
\SetKwFunction{GetPdr}{GetPdr}
\SetKwFunction{ReadPdr}{ReadPdr}
\SetKwFunction{GetBot}{GetBot}
\SetKwFunction{Find}{Find}
\SetKw{Procedure}{Procedure}
\KwIn{$\Psi$, $\alpha_{s}$, $D_{m}$ (neighbor MAC address), $t_{p,R(k)}$;}
\KwOut{$r_{k}$ (selected rate);}
$t_{b,R(k),\Psi}\longleftarrow\infty$,
$\mathbf{E}_{\Psi}\longleftarrow \GetPdr(\Psi)$,
$\mathbf{E^{'}}_{\Psi}\longleftarrow \ReadPdr(\Psi)$\;
\For{$k=(1,\ldots,12)$}{
${E^{''}}_{R(k),\Psi}\longleftarrow\alpha_{s}{E^{'}}_{R(k),\Psi}+(1-\alpha_s){E}_{R(k),\Psi}$\;
$M_{\Psi,k,D_m}\longleftarrow {E^{''}}_{R(k),\Psi}$ \text{(assign PDR to file)}\;
$G_{R(k),\Psi}\longleftarrow\frac{t_{p,R(k)}}{{E^{''}}_{R(k),\Psi}}+\GetBot(E^{''}_{R(k),\Psi})$\;
    \If{$t_{b,R(k),\Psi}>G_{R(k),\Psi}$}{
    $t_{b,R(k),\Psi}\longleftarrow G_{R(k),\Psi}$,
    $r_{k} \longleftarrow k$\;
    }
}
%\BlankLine
%
\Procedure \GetPdr($\Psi$) (read new PDR)\;
\KwIn{$\Psi$, $D_m$;}
\KwOut{$\mathbf{E}_{\Psi}=\{E_{R(1),\Psi},\ldots,E_{R(12),\Psi}\}$;}
\For{$k=(1,\ldots,12)$}{
$E_{R(k),\Psi}\longleftarrow M_{\Psi,k,D_m}$ (read PDR from file)\;
}
%}
%\BlankLine
%
\Procedure \GetBot($E_{R(k),\Psi}$) (get backoff time)\;
\KwIn{$E_{R(k),\Psi}$;}
\KwOut{$t_{b,R(k),\Psi}$;}
$c_m\longleftarrow 31$ (contention window size used),
$t\longleftarrow 0$\;
\For{$i=(0,\ldots,10)$}{
$t\leftarrow t+2^{i}(1-E_{R(k),\Psi})^{i+1}$\;
}
$t_{b,R(k),\Psi}\longleftarrow\frac{c_m}{2}\frac{t+1}{E_{R(k),\Psi}}$ (see~\cite[Eq. 17]{Draves_MobiCom_2004})\;
%\BlankLine
%
\Procedure \ReadPdr($M_{\Psi,k,D_m}$) (PDR from the driver)\;
\KwIn{$M_{\Psi,k,D_m}$ \text{(transmitted and acknowledged packets for $\Psi$, $D_m$ and all $k$)};}
\KwOut{$\mathbf{E}_{\Psi}=\{E_{R(1),\Psi},\ldots,E_{R(12),\Psi}\}$;}
\For{$k=(1,\ldots,12)$}{
$t_x\longleftarrow \text{number of transmitted packets from }M_{\Psi,k,D_m}$\;
$t_{ack}\longleftarrow \text{number of acknowledged packets from }M_{\Psi,k,D_m}$\;
$E_{R(k),\Psi}\longleftarrow \frac{t_{x}}{t_{ack}}$\;
}
\caption{Rate selection}
\label{alg:1}
\end{algorithm}

\section{Proposed LQE: Implementation and Evaluation Environment}
\label{sec:Implementation}

Many previous works demonstrated that current simulation models do not accurately reflect the channel conditions, e.g.~\cite{Newport_Simulation_2007,Lei_ieeewc_2009}. To emphasize on the practicability of the proposed system we have implemented it on real IEEE 802.11b/g hardware and evaluated the network performance on indoor and outdoor testbeds. The results of the evaluation are given in Section~\ref{sec:Evaluation}. But first we need to described in detail the evaluation environment.

\subsection{Indoor Testbed}
\label{sec:indoor_testbed}

\subsubsection{Hardware and Software}
\label{sec:hardware}

The hardware prototype consisted of laptops, each equipped with 3Com OfficeConnect 108Mb 11g PC IEEE 802.11b/g card. The card contains Atheros chip which is supported by the open source Madwifi driver version 0.9.4~\cite{MadWifi}. The laptops ran the GNU/Linux 2.6.24 kernel.

The software implementation architecture is depicted in Fig.~\ref{Fig_implementation}. Ppand~\cite{Jacobsson_phd_thesis}, developed for Personal Networks (PN)~\cite{Niemegeers_WPC2004}, maintains a neighbor list, i.e. creates and splits PN clusters, and combines all the cross-layer information providing an abstraction of multiple link layers to the network layer. As shown in Fig.~\ref{Fig_implementation}, it operates between the network interfaces and the rest of the networking stack. Its purpose is to discover neighbors, handle departures and arrivals of the nodes, authenticate nodes, secure communication, and monitor PDR of the links to those neighbors using methods described in Section~\ref{sec:lqe_proposed}. Ppand generates and processes unicast packets for the purpose of authentication and authorization and and beacon packets for the purpose of link maintenance. Unicast packets are sent by Ppand depending on the network situation, while beacon packets are sent every second. If the beacon has been lost fifty consecutive times then the neighbor is removed from the neighbor list. Packet SNR values were obtained from the driver by Linux's \textit{iwspy} interface~\cite{www:wireless_tools} every time a beacon or data packet was received. To make sure that data traffic does not delay beacon packets, a priority queue was used on each of the interfaces that gives beacons the highest priority. A lost data or beacon packet is assigned the SNR value of the previous packet, while the first non-received packet is assigned the minimum receivable signal strength of --95\,dBm. With every received beacon a new optimal transmit data rate and route is computed. To obtain information on the retransmissions of data packets necessary to evaluate~(\ref{eq:prki}), retrieve the current noise floor (used in computing SNR), and the selected data rate for each node, the Madwifi driver was modified. We note here that the noise floor reported by the Madwifi driver was relatively stable and varied in the range $\approx$\,[--94,\,--98]\,dBm. The transmission power level was set to default 15\,dBm.

In the routing layer an Optimized Link State Routing Protocol (OLSR)~\cite{RFC_OLSR} was implemented using GNU/Linux open source OLSR daemon~\cite{Olsrd} version 4.10.0. A packet delivery ratio aware extension was added to in the routing daemon that made the route decisions based on PDR instead of number of hops. We adapted the routing daemon to get the link quality information from Ppand instead of the daemon's own PDR estimation. The TC message interval was set to 1\,s from 5\,s. The original daemon was designed for the stationary scenario in which link quality does not change much over a 5\,s interval. However, in a mobile scenario, described below, link quality is quite dynamic, so the original interval would limit the potential performance improvement with accurate link quality information. The shorter interval caused very little additional overhead, but resulted in slower route selection decisions. Though the OLSR protocol was used in our testbed, the results obtained in our experiments should apply to other routing protocols as well since the gains were due to improved route selection and not through improved route signaling.

\subsubsection{Measurement Scenarios}
\label{sec:indoor topology}

The proposed link quality aware rate and route adaptation IEEE 802.11b/g ad hoc network has been evaluated for different measurement scenarios using five nodes. We have evaluated the LQE process, rate adaptation and routing independently for different scenarios. All of them are described below.
\begin{figure}
\centering
\subfigure[]{\includegraphics[width=0.49\columnwidth]{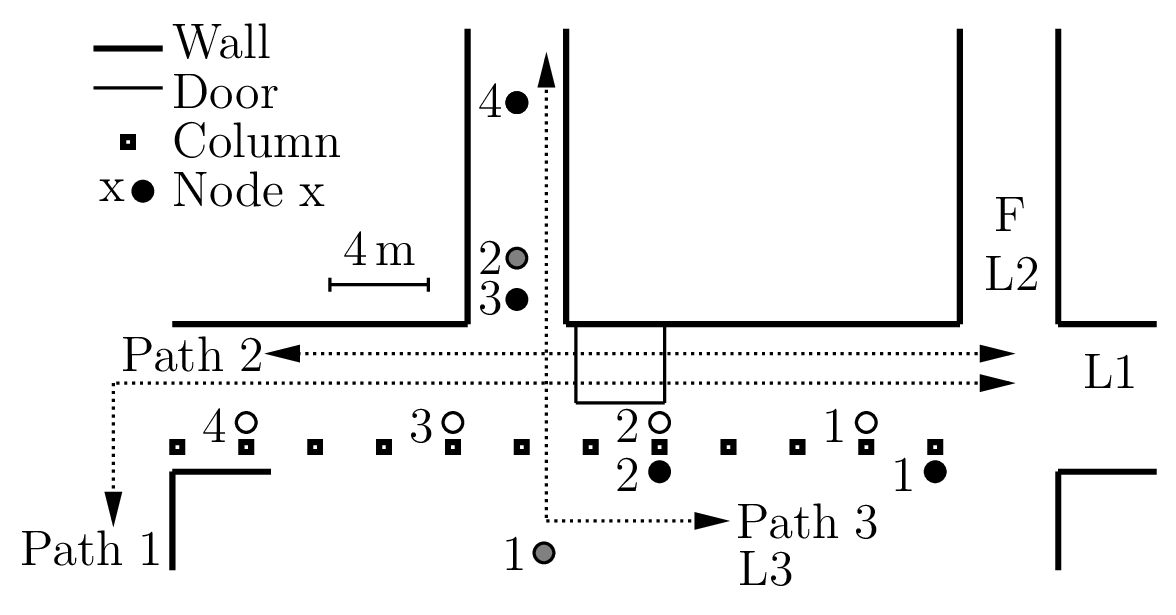}\label{fig:cafeteria}}
\subfigure[]{\includegraphics[width=0.49\columnwidth]{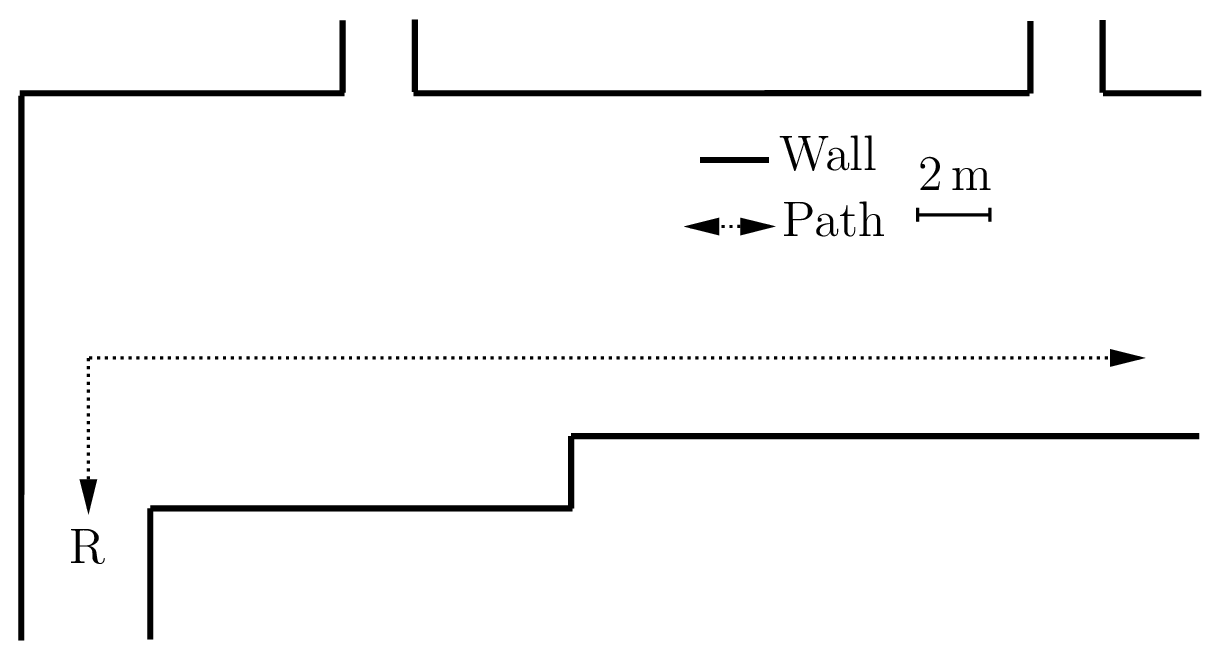}\label{fig:aula}}
\caption{Different topologies used in the evaluation of the proposed network for experiments T1--T6 in (a) cafeteria, and (b) Aula. For more explanation please refer to the text in Section~\ref{sec:indoor topology}.}
\label{fig:topology_indoor}
\end{figure}

In evaluating proposed link quality aware IEEE 802.11b/g network we have set up six different scenarios referred as T1--T6. Scenarios T1--T5 were setup at the cafeteria of Electrical Engineering, Mathematics and Computer Science department of Delft University of Technology. This environment was chosen purposefully, since it contained many walls, pillars, doors, etc., which presents a challenging propagation environment. In all but one scenario, described below with the help of Fig.~\ref{fig:topology_indoor}, one pair communicated on channel 7, which was least affected by other transmissions. Finally, all experiments were performed during late afternoon and night to reduce the probability of interference on that channel. For T6 we set up the experiment in the Aula of Delft University of Technology, which had a different structure, e.g., bigger size, thicker walls and less pillars, than the cafeteria and had more background IEEE 802.11b/g traffic on the selected channel 7. 
\begin{itemize}
\item[T1]
The sender was located at L1 while the mobile node moved along path 1 with the speed of $\approx$\,1\,m/s. This scenario was needed to evaluate the proposed method in a mobile environment and to observe the SNR mapping table for a broad range of SNR values. For 80\% of path 1 the link had LOS and for the remaining 20\% was shadowed by the wall. The shadowed portion of the path was needed to obtain very low values of SNR. In LOS portion we could not obtain a SNR lower than 15\,dB. 

\item[T2]
The sender was located at L1 while the receiver moved along path 2, in stop-and-go fashion, i.e., the receiver resides in one location for 1\,min and then moves 1\,m away from the sender (in the case of LQE measurements), and waits 3\,min, then moves 4\,m away from the sender (in the case of rate adaptation measurements). The whole process took about 50\,min. 

\item[T3]
The same as T2, except the sender moved along path 2 and the receiver was located at L2.

\item[T4]
The same as T3, with a third node located at F separated 20\,cm from L2 and not involved in the same network. Node F broadcasted a large amount of traffic, i.e., 500\,packets/s, at the same data rate and packet size as the sender at L1, but on a channel two away from that of the sender and receiver. This scenario served the purpose of evaluating the impact of adjacent channel interference on the proposed LQE method.

\item[T5]
The same as T4, except the interfering node transmitted packets on the same channel as the sender and receiver. This scenario served the purpose of evaluating the impact of co-channel interference on the proposed LQE method.

\item[T6]
The topology for this rate adaptation experiment is the same as in T1, compare Fig.~\ref{fig:cafeteria} and Fig.~\ref{fig:aula}. The nodes had SNR profiles created in the cafeteria. This experiment was set up to show the ease of adaptability of the proposed LQE to different types of environments and to evaluate the process given by (\ref{eq:e_sj}).
\end{itemize}

Finally, in case of routing process evaluation we have set up the following three scenarios, called as R1--R3.

\begin{itemize}

\item[R1]

The sender and receiver were located close to each other at L3, see Fig.~\ref{fig:cafeteria}. The sender followed path 3 away from receiver and returned. For this path two nodes, marked as grey circles in Fig.~\ref{fig:topology_indoor}, forwarded the traffic. The movement was repeated 10 times and lasted 90\,s. Before each movement along path 3 the sender waited 30\,s at L3, to let the rate adaptation mechanism stabilize to de-correlate each movement from the previous.

\item[R2]

The sender located at L1, see Fig.~\ref{fig:topology_indoor}, sent traffic to each of its four neighbors one by one, marked as white circles in Fig.~\ref{fig:topology_indoor}.  The sender send a constant flow of UDP traffic for 5\,min to one receiver before it targeted the next node. Nodes were allowed to forward the traffic. 

\item[R3]

The same as R2, but the receiving nodes were located differently, marked as the black circles in Fig.~\ref{fig:topology_indoor}. This was the non-LOS communication scenario in a ad hoc network. 
\end{itemize}

It is important to discuss why we decided to limit ourselves to the line (tandem) topology with five nodes. First, in real life, large scale ad hoc networks, where dozens of nodes are actively transmitting is rare. Second, due to the IEEE 802.11 transmission medium characteristics end-to-end throughput decreases dramatically with increasing hop count~\cite{Li_Ciit_2004}, especially for very high data rates, i.e. $R\gtrsim$\,11\,Mbps. Thus for many hops (greater than five) random routing and contention effects will invalidate the throughput comparison of different methods.

\subsection{Outdoor Testbed}
\label{sec:outdoor_testbed}

\subsubsection{Hardware and Software}
\label{sec:hardware_outdoor}

All nodes used in the outdoor experiments were running IEEE 802.11b/g. Each device was equipped with two radios using the Atheros chipset, the same as used in the indoor testbed described in Section~\ref{sec:indoor_testbed}. The operating system was a modified GNU/Linux kernel version 2.6.28 using the modified Madwifi driver version 0.9.4. Because no rate and route adaptation experiments were performed outdoors, Ppand was not needed. Instead a socket-based application that created SNR profiles was developed based on Madwfi. Note, that just like in the indoor experiments noise floor oscillated around --95\,dBm.

\subsubsection{Measurement Scenarios}
\label{sec:scenarios_outdoor}

The outdoor experiments were performed on the Quail Ridge Wireless Mesh Network~\cite{wu_tridentcom_2007}. The network was located in Napa County, California, on a southern peninsula of Lake Berryessa and consisted of hilly and densely forested terrain. There were 34 nodes at the time of testing, all placed at varying elevations due to the terrain spanning 2000 acres. Distances between nodes ranged from 1 mile to a few hundred meters. Directional antennas were used for point-to-point links on top of the hills, while omnidirectional antennas were used for lower elevations. The following experiment was performed.

\begin{itemize}

\item[T7] A randomly selected sender sent a data to a randomly selected receiver for about 30 minutes in all available data rates. This experiment was needed to obtain an SNR profile for outdoor scenario.

\end{itemize}

\section{Evaluation of the Proposed LQE and its Usage in Rate and Route Adaptation: Accuracy, Efficiency and Adaptability}
\label{sec:Evaluation}

Given the explanations of the proposed method in Section~\ref{sec:Proposed_system} and the implementation and measurement scenarios in Section~\ref{sec:Implementation} we present the results of our measurements. We start with the discussion on the proposed LQE procedure focusing on SNR profile-based LQE.

\subsection{Properties of SNR Profile}
\label{sec:lqe_evaluation}

\subsubsection{Data Rate versus SNR Profile}

\begin{figure}
\centering
\subfigure[]{\includegraphics[width=0.32\columnwidth]{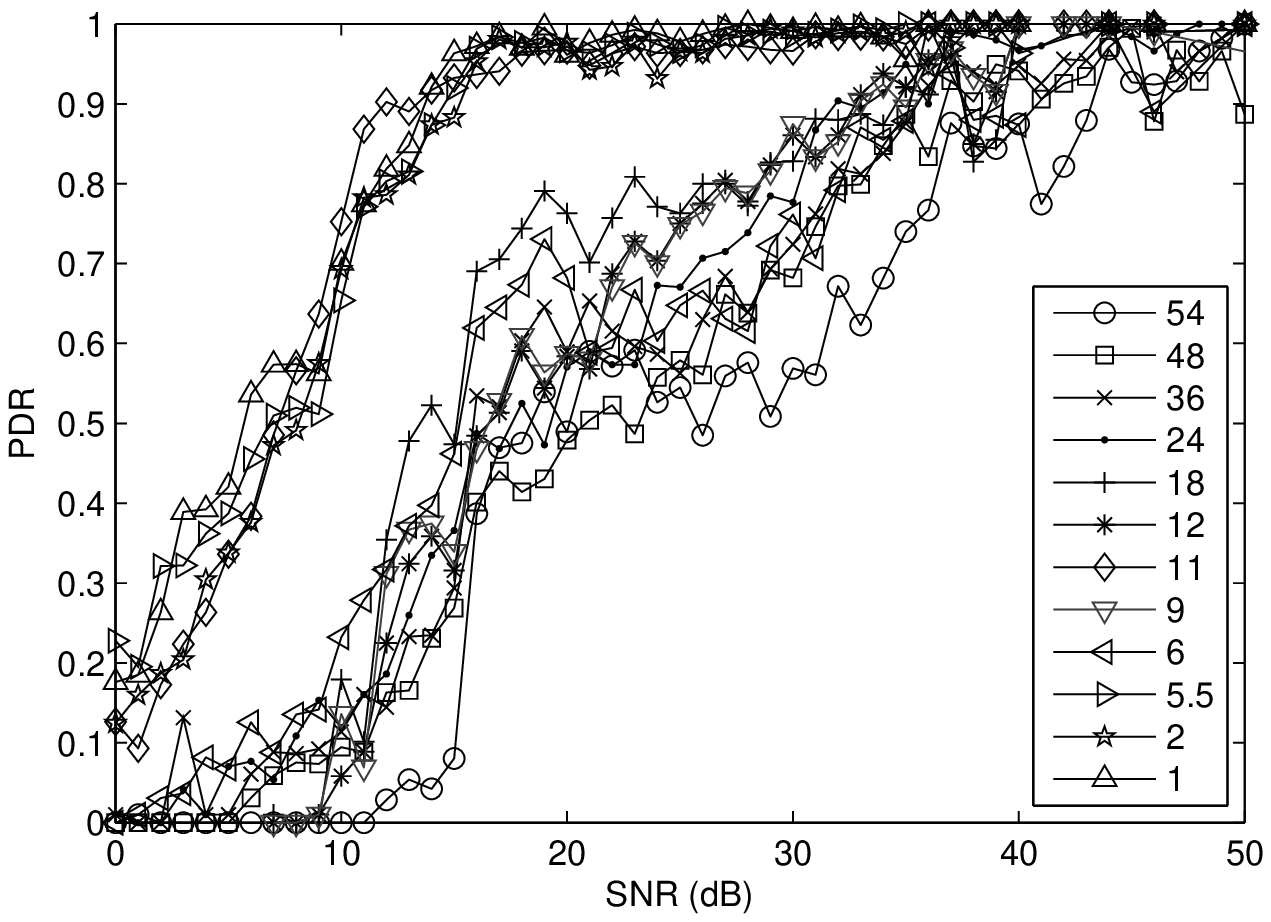}\label{Fig_rate}}
\subfigure[]{\includegraphics[width=0.32\columnwidth]{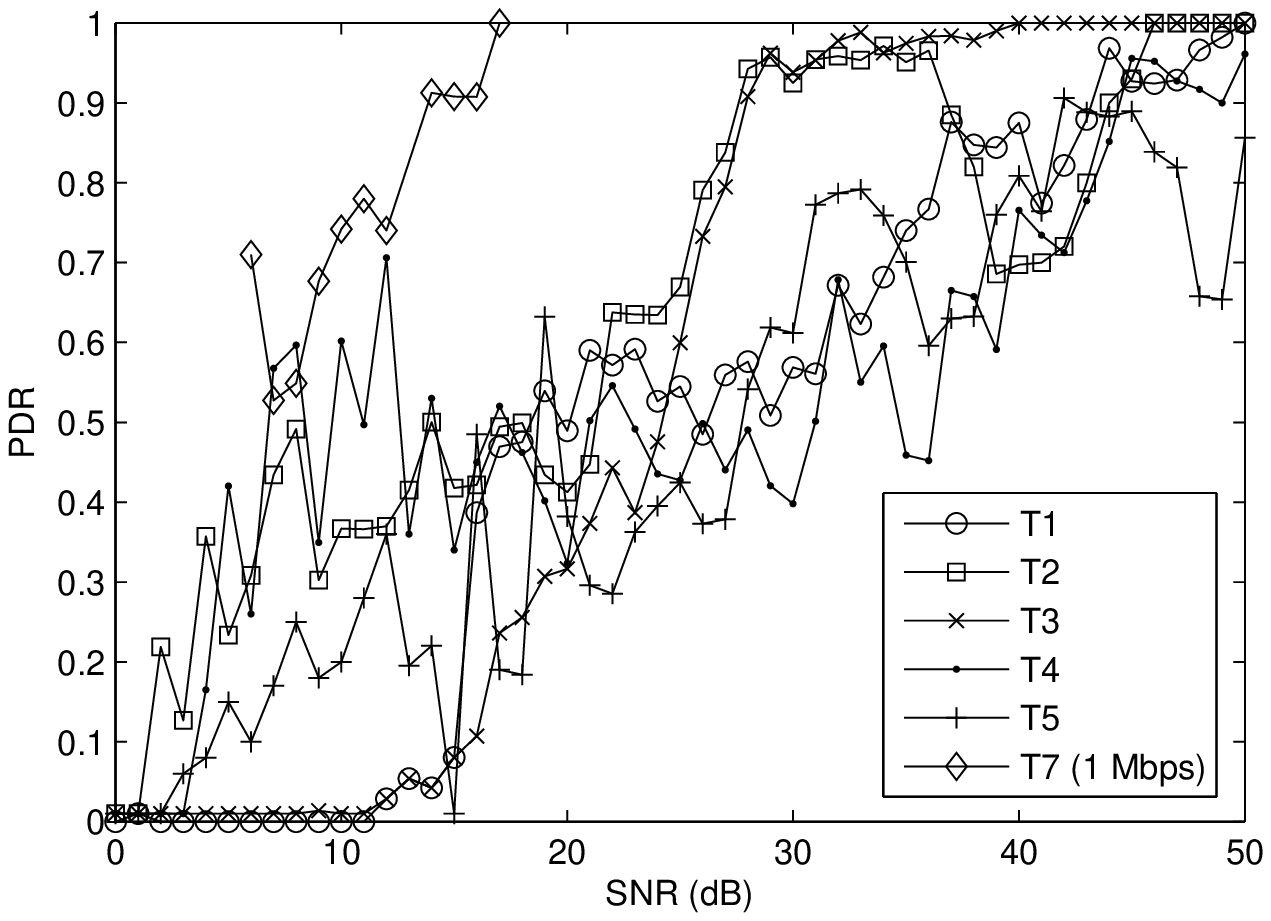}\label{Fig_Stationary_factors}}
\caption{Examples of SNR profiles: (a) profiles for all data rates of IEEE 802.11b/g in T1, and (b) SNR profile for 54\,Mbps data rate in T1 and for 1\,Mbps in T7 and different measurement scenarios, as described in Section~\ref{sec:indoor topology} and in Section~\ref{sec:scenarios_outdoor}.}
\end{figure}

We have obtained SNR maps for all IEEE 802.11b/g data rates in scenario T1 and shown the results in Fig.~\ref{Fig_rate}. We can see that there is a clear difference between IEEE 802.11 b and g data rates. However, it is not clear that for a certain SNR, a higher data rate will result in lower packet deliver rate.

The IEEE 802.11b rates have higher PDR even at lower SNR than IEEE 802.11g rates. This is due to the modulation used by each system. IEEE 802.11b uses DSSS and CCK whereas 802.11g uses OFDM. This results in two distinct groups of curves visible in Fig.~\ref{Fig_rate}.

The SNR maps of IEEE 802.11g rates show more fluctuations, which can be attributed to propagation environment, narrowband interference, ISI and vulnerability of modulation to bit errors\footnote{We note that even with minimized interference on the considered measurement channel and very long measurement times, i.e. longer that ten hours, we were not able to reproduce such smooth SNR profiles as in~\cite[Fig. 2, Fig. 5]{Zhang_Infocom_2008}.}. To achieve higher bit-rates, IEEE 802.11g uses BPSK, QPSK, 16-QAM, and 64-QAM with different code rates. Higher rates have higher vulnerabilities to bit errors than lower rates. Hence, higher SNR is required by IEEE 802.11g rates to achieve a PDR of 1. From Fig.~\ref{Fig_rate} it can be seen that a SNR $>$ 40\,dB is required to achieve a PDR $\geq$ 0.9 reliably by 64-QAM and code-rate $3/4$ (54\,Mbps).

\subsubsection{Environment versus SNR Profile}
\label{sec:env_vs_snr}

Apart from data rate, different environments influence the SNR profile. We have created SNR profiles for 54\,Mbps data rate from experiments in scenarios T1--T5 and T7, and present our results in Fig.~\ref{Fig_Stationary_factors}. The received SNR is higher for LOS communication (T1 and T2) and hence the PDR is higher. However in T1 the sender moves continuously and the decoding of the signal depends on the propagation environment causing fluctuations in the graph. Similar arguments follow for other scenarios.

The SNR profiles of T2 and T3 (stationary sender in quasi-LOS and non-LOS, respectively) are different in the low SNR region, i.e. $\approx$\,[0,\,25]\,dB and similar in the high SNR region. The PDR from T1 is not perfect even at high SNR and is much lower than that from scenario T2 and T3. For T4 interference caused by the independent transmitter increased the noise level on the desired channel. Finally, for T5 we can see that for high SNR PDR is not as high as from the pure non-LOS stationary scenario T3. For low SNR values, due to large variance of the channel, the PDR varies great and sometimes can achieve high PDR.

In T7, due to the outdoor testbed constraint, we could not obtain mapping at high data rates, including 54\,Mbps, since SNR was too low for high data rate packets to go through the selected link. To show how the mapping looks like in outdoor links, we plot mapping relation for 1\,Mbps instead. Moreover, for the outdoor testbed, since nodes are fixed and largely separated, the channel is far more stable than indoor channels than seen in T1-T6, so single link cannot provide a wide range of SNRs as seen in indoor. Therefore we combined SNR maps from three links to obtain the mapping.

\subsection{Accuracy of Proposed LQE}
\label{sec:LQE accuracy}

To compare the accuracy of SNR profile-based and packet counting-based LQE, we use data packet counting method as our benchmark, which was used in~\cite{zhang_2009_tmc}. For a given data rate and scenario we compute the estimation error using the mean absolute error metric, used widely in evaluating time series forecasting accuracy, as
\begin{equation}
\Delta=\frac{1}{m}\sum_{i=1}^{m}|E_{R(k),i,d}-E_{R(k),i}|,
\label{Eq_accuracy calculation}
\end{equation}
where $m$ is the total number of collected packet sampling intervals, which contains $E_{R(k),i}$, both for SNR profile estimation and broadcast packet based estimation, beacon based LQE, and $E_{R(k),i,d}$, which is LQE obtained using data packet counting.

Our results are split in two parts. First we analyze a static profile where one SNR profile is used in all measurement scenarios. Later we analyze the on-line updated SNR profile.

\subsubsection{Static SNR Profile}
\label{sec:result_static_profile}

\begin{figure}
\centering
\subfigure[]{\includegraphics[width=0.32\columnwidth]{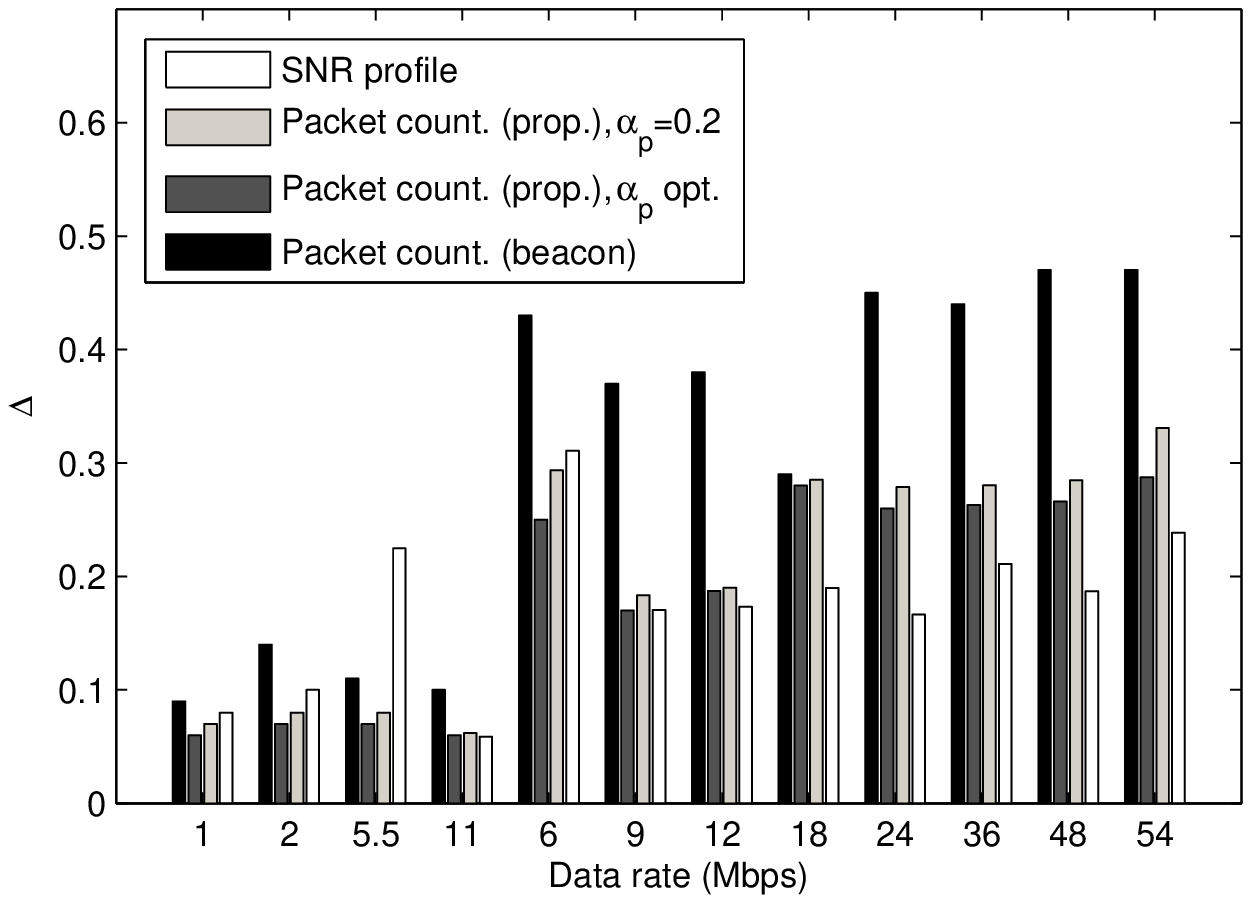}\label{Fig_LQE_error_mobile}}
\subfigure[]{\includegraphics[width=0.32\columnwidth]{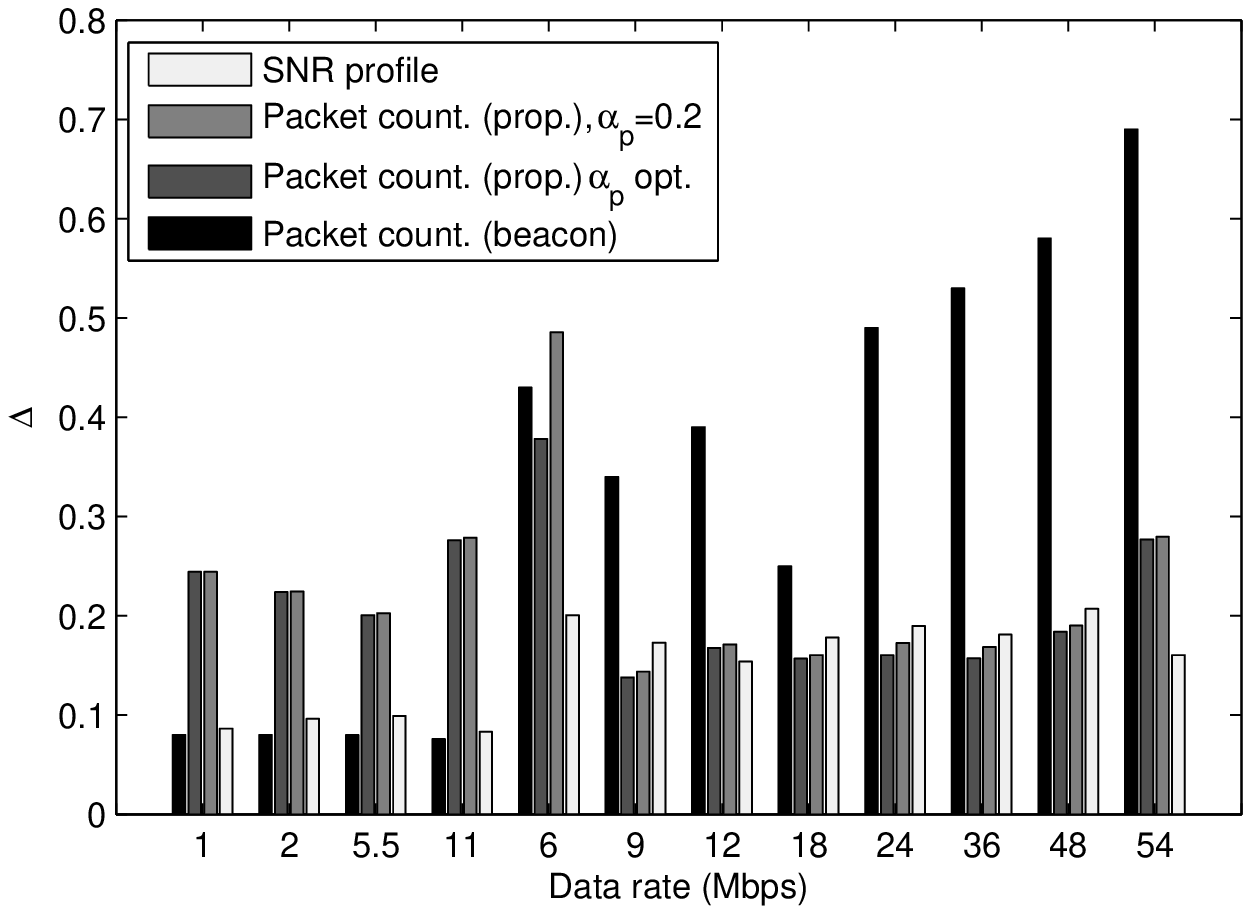}\label{Fig_LQE_error_stationay}}
\subfigure[]{\includegraphics[width=0.32\columnwidth]{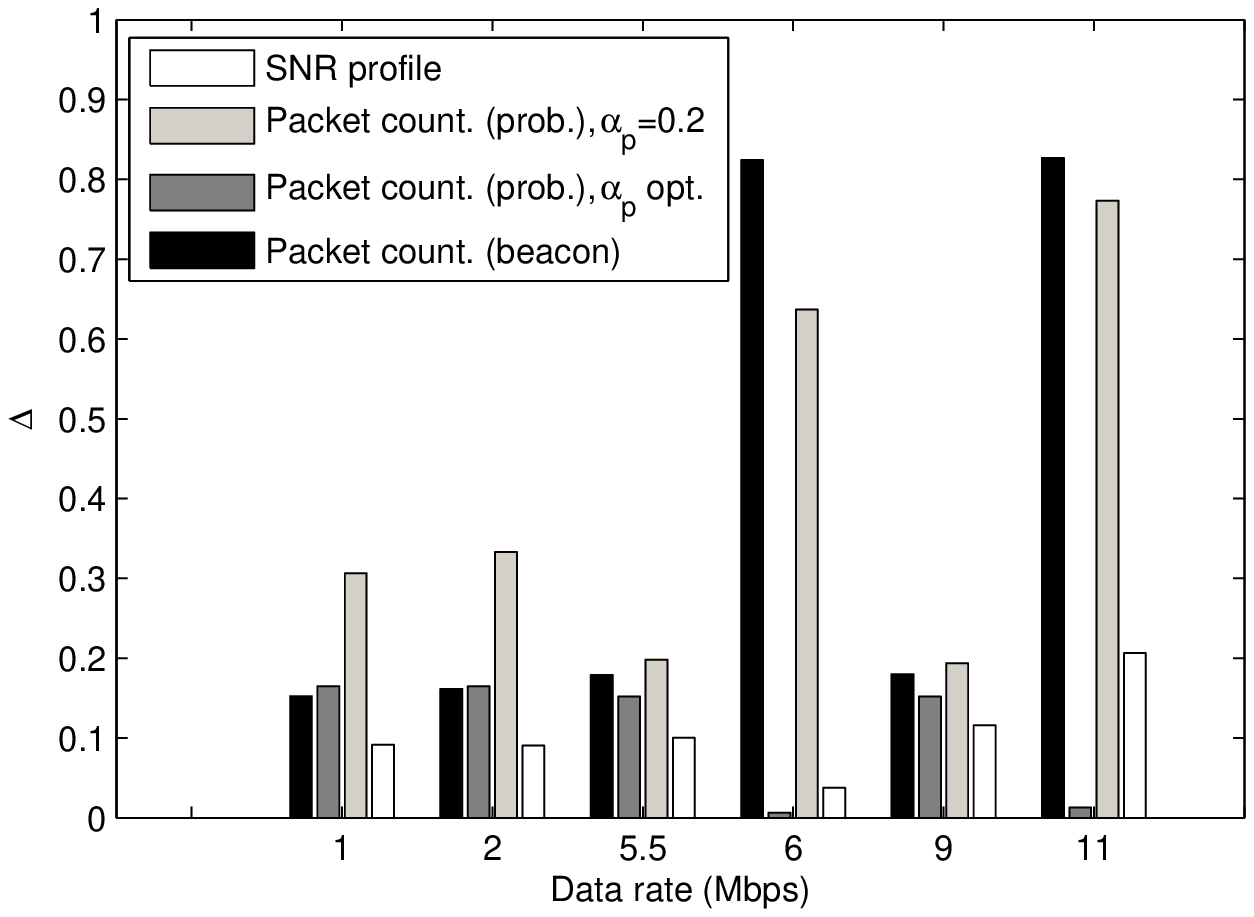}\label{Fig_LQE_error_outdoor}}
\caption{LQE error $\Delta$ given by (\ref{Eq_accuracy calculation}) in (a) T1 scenario, (b) T2 scenario and (c) T7 scenario. Data rates of $R$ are sorted in the figure such that first left four are IEEE 802.11b data rates and the remaining ones are IEEE 802.11g.}
\end{figure}

Here we focus only on scenario T1, T2 and T7. The experimental
results for the T1,T2 and T7 scenarios are shown in
Fig.~\ref{Fig_LQE_error_mobile}, Fig.~\ref{Fig_LQE_error_stationay},
and Fig.~\ref{Fig_LQE_error_outdoor}, respectively.

In case of proposed broadcast packet based LQE we have separately measured link quality with $\alpha_p=0.2$, which is an arbitrary yet reasonable value for the EWMA filter, and for the optimal value of $\alpha_p$, which gives the lowest $\Delta$.

In T1 the beacon based estimation has the highest error for all data rates except 5.5\,Mbps. In T2 the method under performs for all IEEE 802.11g data rates in T2, mostly because difference between probing and data packet and higher channel variance. For the proposed broadcast packet based probing the error is much less because the probing packet is the same as the data packet. Still, SNR profile based LQE for the majority of rates in T1 (except for 1, 2 and 5.5\,Mbps) outperforms the two methods. For T2 it achieves comparable performance as the proposed broadcast based packet counting. Interesting conclusions can be drawn for low data rates, primarily over IEEE 802.11b, in scenario T2. Since these data rates have much larger transmission range and evidently more robust modulation schemes compared to IEEE 802.11g, beacon based LQE has similar or less estimation error than SNR profile. In both environments, optimized broadcast packet based LQE has almost the same performance as with $\alpha_p=0.2$, which proves that the chosen smoothing factor is sufficient.

Using the same processing methods for Fig.~\ref{Fig_LQE_error_mobile} and Fig.~\ref{Fig_LQE_error_stationay}, we plot estimation error $\delta$ for different dates rate obtained in T7. Due to the phenomena described in Section~\ref{sec:env_vs_snr}, we use only eight lower data rates to calculate LQE accuracy in outdoor links, instead of all twelve. The conclusions are in line with the ones obtained for T1 and T2. The SNR mapping is the best LQE method, being worse only for optimal packet counting-based LQE for 6\,Mbps and 11\,Mbps. This can be explained as follows. Because for the selected link in T7, the number of data that successfully passed through was very low, so if we start with PDR of zero in estimation and the optimal $\alpha_p$ is zero, then estimated PDR is good, since it also takes zero for the estimation. Needless to say, the selection of optimal $\alpha_p$ is just a benchmark instead of real practical LQE method, since in reality the optimal $\alpha_p$ can never be used due to huge number of neighbors and dynamic links. Which means that if we use optimal $\alpha_p$,  computation for the optimal value has to be performed for each neighbor frequently, and not all nodes have enough processing power to do so. Finally, this method in most scenarios still performs worse than SNR mapping, as seen in Fig.~\ref{Fig_LQE_error_mobile} and Fig.~\ref{Fig_LQE_error_stationay}.

It is important to note that since for the IEEE 802.11g data rates data packets are rarely transmitted, most of the time the PDR estimation for those data rates is almost zero, but beacon still can pass through easily, so that beacon packet-based LQE made huge error when it tried to estimate the IEEE 802.11g data rates. The error for the broadcast packet-based LQE is due to the fact that broadcast packets are not retransmitted while the unicast packet are. The optimal $\alpha_p$ could bring very accurate LQE, however, as it is seen in Fig.~\ref{Fig_LQE_error_mobile} and Fig.~\ref{Fig_LQE_error_stationay}, this method can never be used in reality.

Based on the results from the above three scenarios, we conclude that the SNR profile based or broadcast packet counting based methods perform better than methods based on beacon packet counting. In some scenarios, the proposed broadcast based method performs slightly better than the SNR based method. However, due to the facts described in Section~\ref{sec:LQE_packet} probing all data rates and packet sizes will cause huge overhead for the network. Therefore, the SNR profile based method is the only method that maintains both accuracy and efficiency for all scenarios.

\subsubsection{Dynamic SNR profile: Mitigating the Need for Context Awareness}
\label{sec:result_self_updating}

As we have discussed in Section~\ref{sec:context} it is possible to mitigate the need for a context aware system to obtain the required accuracy in a certain scenario by optimizing parameters of the SNR profile based LQE as the data traffic passes through the node.

\begin{figure}
\centering
\includegraphics[width=0.4\columnwidth]{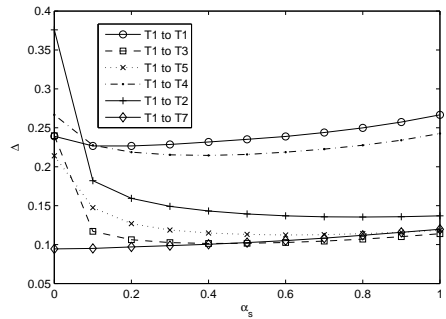}
\caption{Estimation error $\Delta$ for all values of $\alpha_s$ and changes in measurement environments.}
\label{Fig_Updating effect}
\end{figure}

We will use the SNR profile generated in T1 at 54\,Mbps in a different scenario. The results for the other data rates will behave similarly. Scenario T1 is used as the initial one, and with SNR profile-based LQE, see Section~\ref{sec:snr_profile}, the speed of the update in new environment is controlled by $\alpha_s$. The result is presented in Fig.~\ref{Fig_Updating effect}. We conclude that in all scenarios, indoor and outdoor, updating the SNR profile at the optimal rate achieves better estimation than the static profile, i.e. static profile (when $\alpha_s=0$).

Although for the outdoor scenario $\delta$ is the smallest when the $\alpha_s$ is zero, however $\delta$ is very small for all $\alpha_s$ which still shows that our method can be effectively used in different scenarios.

\subsection{Evaluation of the Responsiveness of LQE-Based Rate Adaptation}
\label{sec:result_rate_adaptation}

We now look at the adaptability of the proposed LQE process, from the perspective of rate adaptation and focus on SNR profile based LQE only. Just like in Section~\ref{sec:LQE accuracy}, we split the discussion focusing first on static SNR profile and later on its updating.

\subsubsection{Static SNR Profile}
\label{sec:rate_static_profile}

\begin{figure}
\centering
\subfigure[]{\includegraphics[width=0.32\columnwidth]{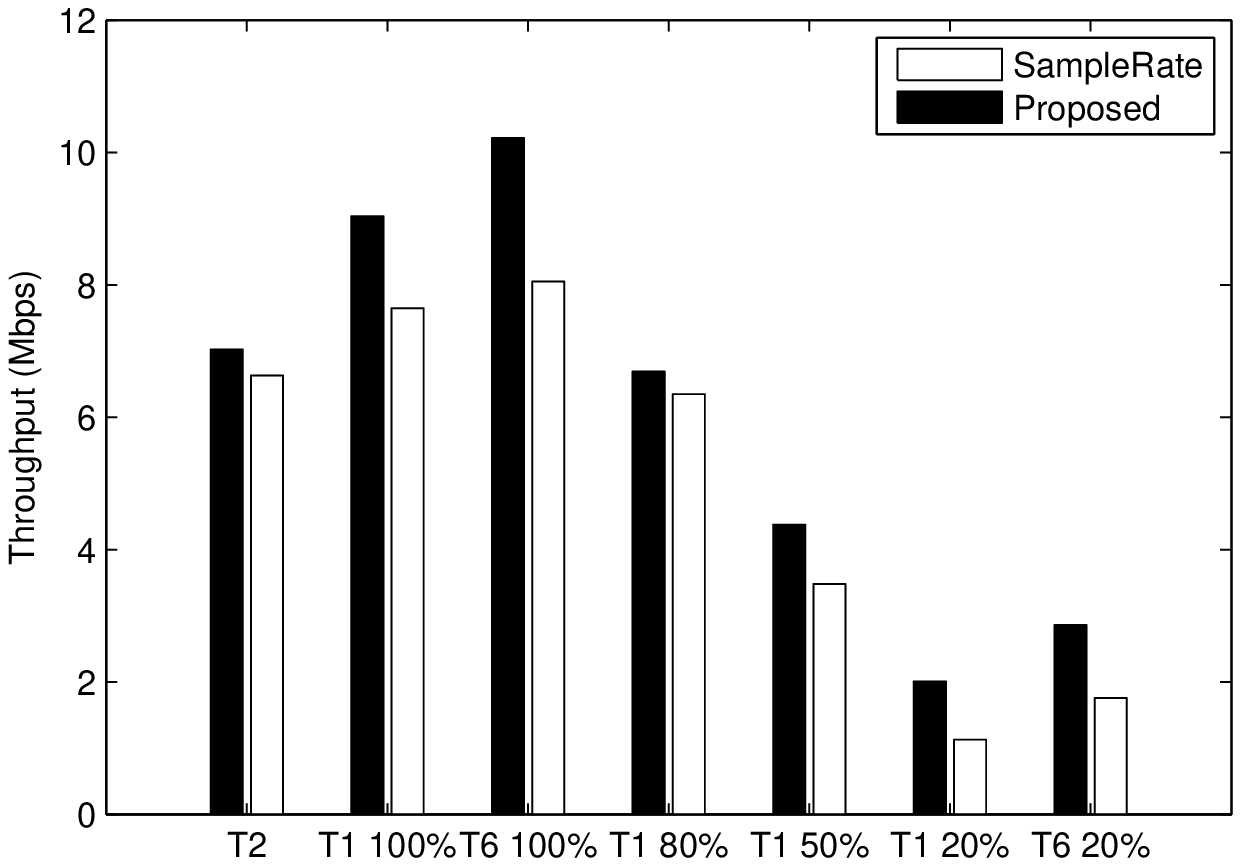}\label{Fig_Rate_throughput_compare}}
\subfigure[]{\includegraphics[width=0.32\columnwidth]{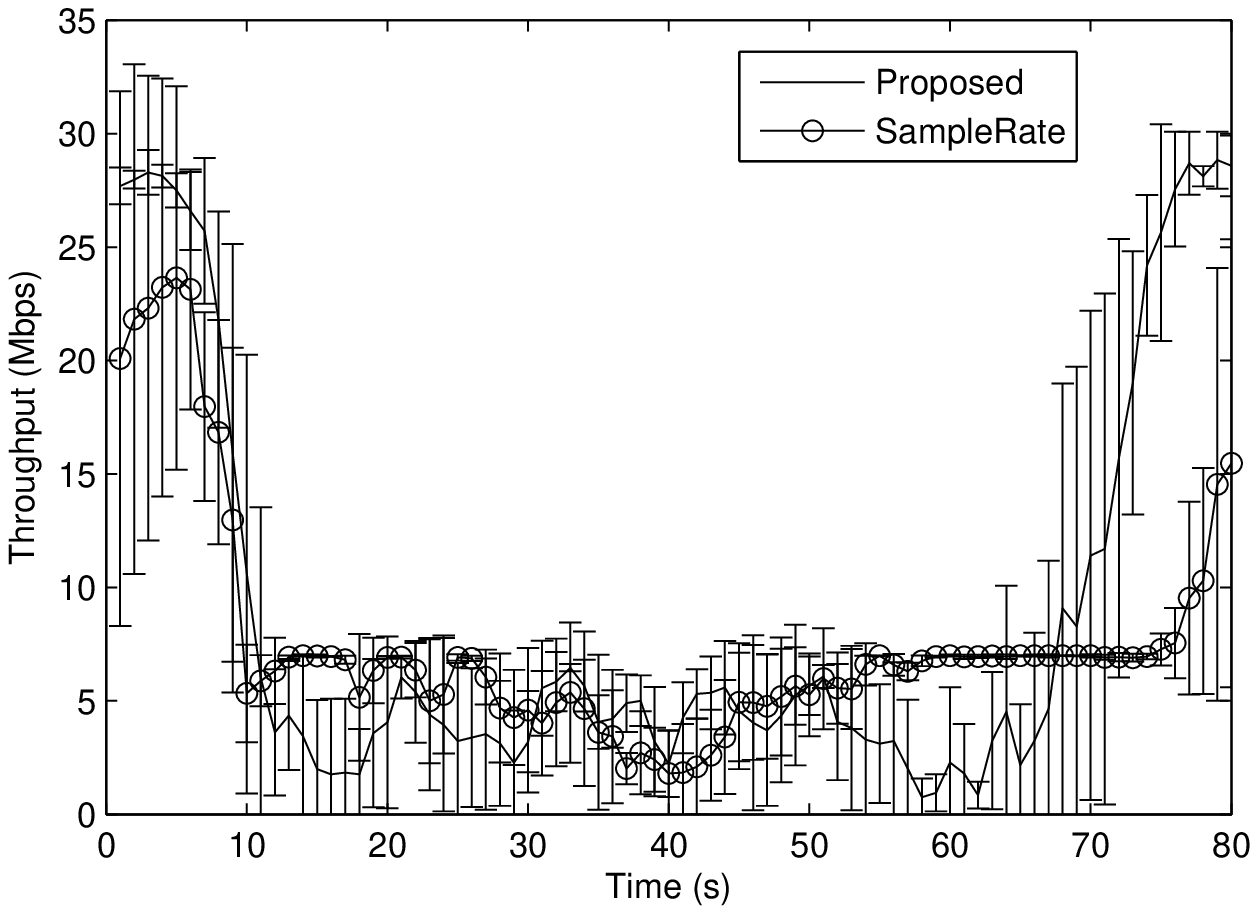}\label{Fig_Rate_throughput_instantaneous}}
\caption{The performance of proposed rate adaptation mechanism: (a) in comparison to SampleRate in T1 and T2 scenarios (percentage values represent average amount of traffic flowing between sender and receiver); and (b) time evolution in relation to SampleRate in T1 scenario.}
\end{figure}

We test the performance of the proposed rate adaptation mechanism in two scenarios, T1 and T2, and compare it against the SampleRate protocol, where in contrary to our approach a constant traffic flow was used for rate adaptation. For T1 we have tested how the amount of data traffic affects the rate adaptation efficiency. In four different experiments we varied the amount of traffic flowing between nodes over period of 400\,s such that, on average, the network was loaded \{100, 80, 50, 20\}\% of the time. The results are presented in Fig.~\ref{Fig_Rate_throughput_compare}. We can see that our proposed rate adaptation method outperforms the SampleRate mechanism, especially in low data traffic scenarios (almost two times the throughput of SampleRate with 20\% loading). In the T2 scenario (stationary, LOS) the proposed rate adaptation mechanism is again better than SampleRate, in this case by 0.5\,Mbps. For T6 we clearly see that, starting with the SNR profile created in the cafeteria, nodes were able to update the profile and increase throughput by $\approx$\,23\% for 100\% loading and $\approx$\,33\% for 20\% loading. This proves the statement that initial SNR profile can be created only once in the entire lifetime of the device, see Section~\ref{sec:snr_profile}, allowing for easy update every time it operates in a new environment.

In Fig.~\ref{Fig_Rate_throughput_instantaneous} we plot the instantaneous throughput versus time for the T1 scenario. We observe that when the transmitter is closest and farthest away from the receiver, the proposed rate adaptation mechanism is better than Sample Rate. However, in the middle of the range the performance is not significantly better, in fact there are cases when Sample Rate performs better, see Fig.~\ref{Fig_Rate_throughput_instantaneous} for [50,\,65]\,s. This is because during the experiment the receiver, passing through the corridor, received packets with SNR in the range of 9 to 15\,dB.  Within this range our algorithm switches often between 802.11b and g rates, whichever has the lowest total instantaneous transmission time. This causes our algorithm to achieve a range of throughput values. SampleRate however plays safe by sticking to rate of 11\,Mbps, and achieves a constant throughput.

\subsubsection{Updating Process in Rate Adaptation}
\label{sec:rate_updating_based}

It is important to evaluate the impact of $\alpha_s$ on the rate adaptation process. In scenario T1 we use a different $\alpha_s$ each experiment and measure the achieved throughput,  repeating the experiment for one $\alpha_s$ in the same scenario at three different times. In Table~\ref{Tab_alpha_value} we use the same $\alpha_s$ as in Fig.~\ref{Fig_Updating effect}, which is used for updating the SNR profile based on new data traffic. However, it is important to note that the optimal $\alpha_s$ is not necessarily the same in these two cases, since in Fig.~\ref{Fig_Updating effect} we only update the SNR profile for one data rate, and in Table~\ref{Tab_alpha_value} we update the SNR profile for several data rates. Nevertheless, we can see that if $\alpha_s>0$ the performance is quite similar in both cases for different $\alpha_s$ values.

\begin{table}
\centering
\caption{The proposed rate adaptation performance for different $\alpha_s$ values}
\label{Tab_alpha_value}
\begin{tabular}{c|c}
\hline
$\alpha_s$ & Throughput [Mbps]\\
\hline\hline
0 & [6.80, 8.55]\\
0.05 & [8.30, 9.63]\\
0.1 & [8.77, 9.59]\\
0.2 & [8.20, 9.48]\\
0.3 & [7.64, 9.11]\\
0.5 & [8.14, 8.67]\\
0.9 & [7.63, 8.39]\\
\hline
\end{tabular}
\end{table}

\subsection{Evaluation of the Efficient LQE-Based Routing}
\label{sec:result_routing_improvement}

\begin{figure}
\centering
\includegraphics[width=0.32\columnwidth]{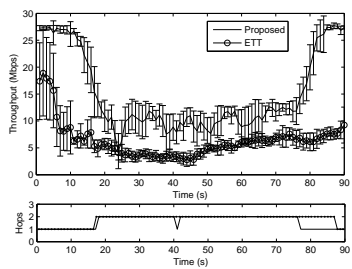}
\caption{The OLSR throughput obtained for scenario R1 with proposed system and ETT metric. Please note missing information on hop count starting at 61\,s, which is due to TC message being lost.}
\label{Fig_result_route_mobile_instan}
\end{figure}

Experiments at the routing layer are needed to evaluate the overall
efficiency of a link quality aware system. First in
Fig.~\ref{Fig_result_route_mobile_instan} we plot the throughput
versus time for the rate adaptive scenario R1.

We can see that our proposed mechanism can outperform the current OLSR implementation using ETT. In the initial phase our mechanism tends to select better data rates which achieve greater single hop throughput. We can see from the hop chart (bottom of Fig.~\ref{Fig_result_route_mobile_instan}) that after 20\,s both mechanisms start to use the two-hop route. As observed previously, due to better selection of data rate the throughput with our method is also much higher than the packet counting based system. This can be seen as the transmitter moves towards the receiver (between 60\,s and 80\,s). This is due to the slow reaction time of SampleRate and the inaccurate LQE of ETT. The packet counting based system still chose a 2-hop route even when the distance between the transmitter and receiver is less than that between the transmitter and intermediate node. This results in much worse performance than our proposed link quality aware system.

\begin{figure}
\centering
\subfigure[]{\includegraphics[width=0.32\columnwidth]{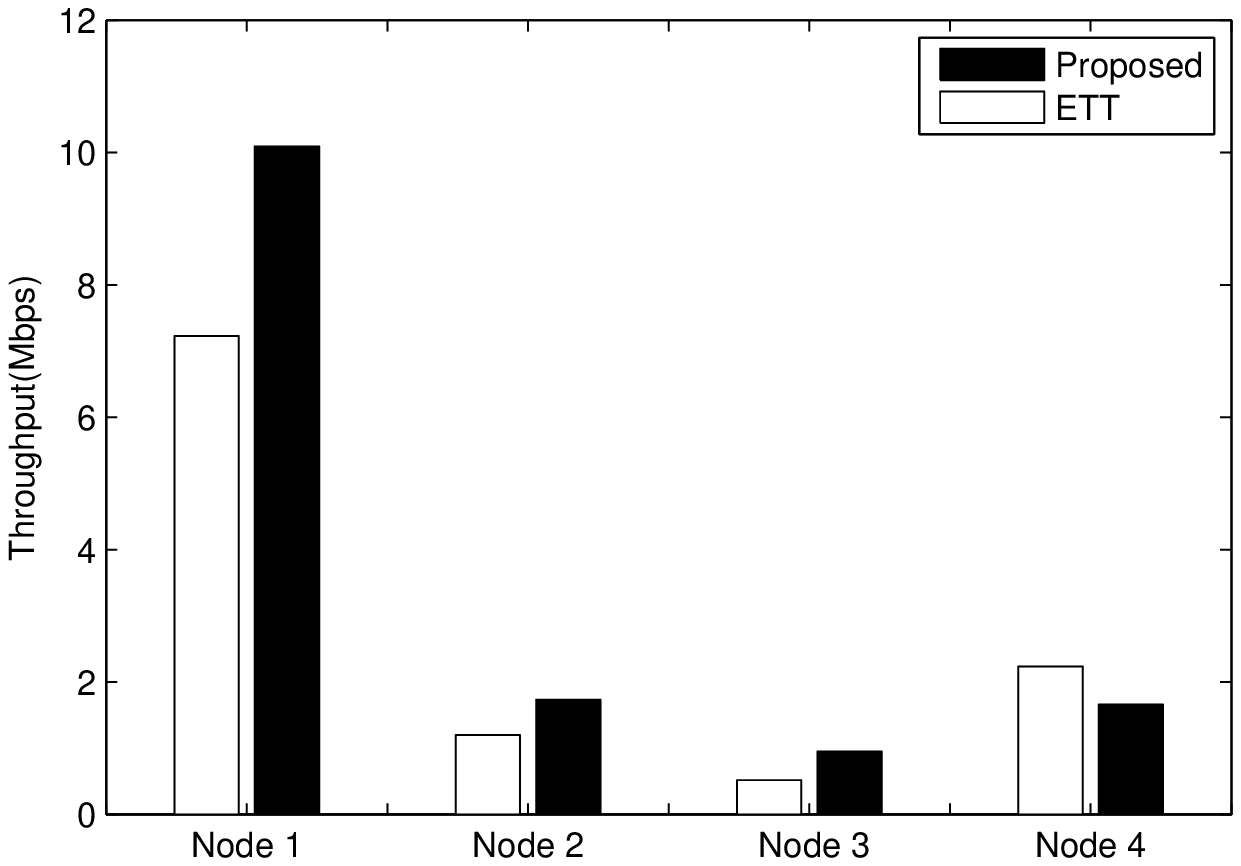}\label{Fig_result_route_stationary_3pillar}}
\subfigure[]{\includegraphics[width=0.32\columnwidth]{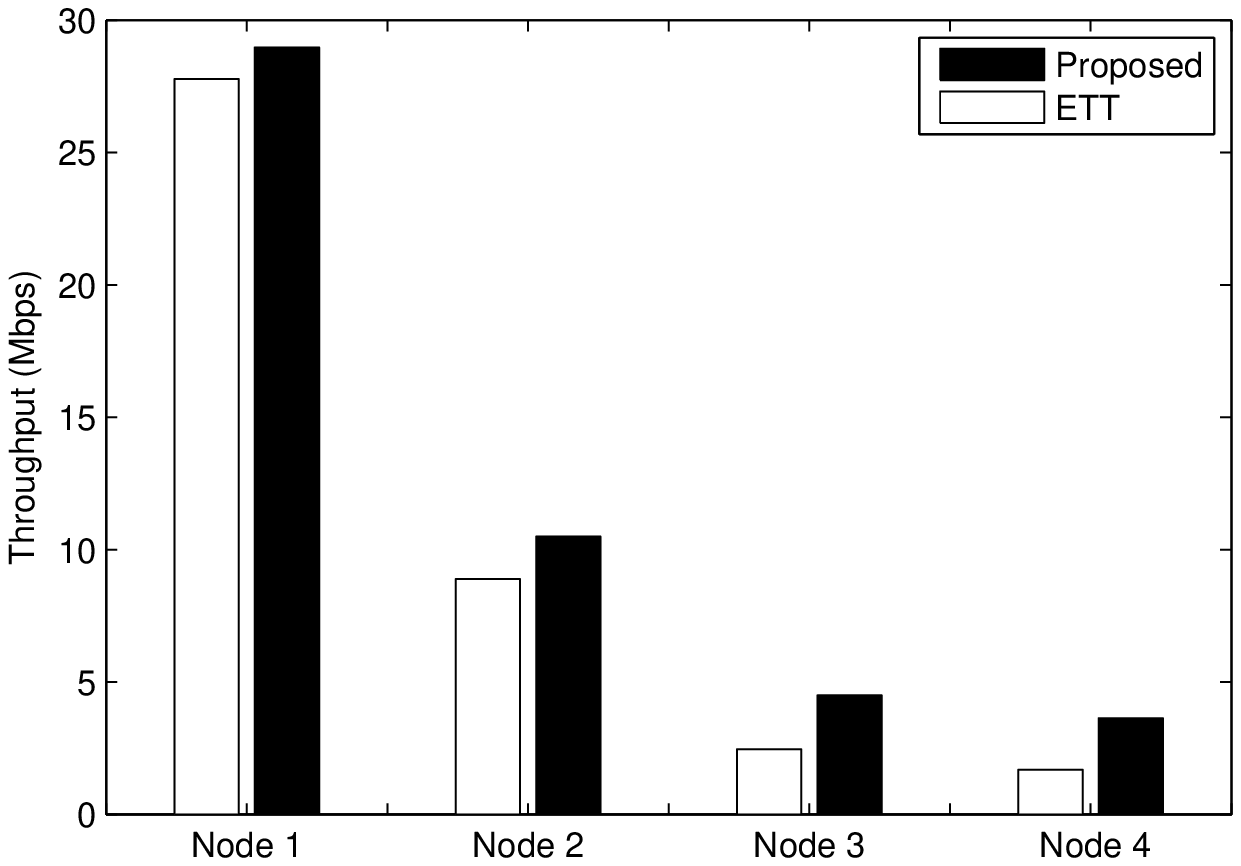}\label{Fig_result_route_stationary_4pillar}}
\caption{The throughput for each receiver in topology (a) R2 and (b) R3.}
\end{figure}

The results of experiments in R2 are given in Fig.~\ref{Fig_result_route_stationary_3pillar}. We can see that in this situation our proposed mechanism performs better than OLSR with ETT. When the receiver was node 2 or node 3, only one or two hops were traversed. Due to the fact that the connection between node 1 and the sender is not satisfactory, either a one or two hop route between the receiver and sender will result in decreased throughput. An interesting behavior is seen for node 4 which is supposed to achieve the lowest throughput due to its greatest distance from the sender. In this case the SampleRate-based OLSR performs better than our proposed system while both of them outperform node 3 and node 2 in terms of throughput. This is presumably due to sudden link breaks that decreased temporarily the throughput to zero, which resulted in fixed routing outperforming our proposed LQE-based routing.

The results of experiments for the scenario R3 are shown in Fig.~\ref{Fig_result_route_stationary_4pillar}. Similar to the results in R2, the first node uses a one hop route to the sender and the performance of our algorithm is a little better than the classic OLSR due to the same reason as described for R2. However, when the receiver moves further away from the sender (nodes 2, 3, and 4, respectively), our method performs even better and the throughput is much higher than in the SampleRate based OLSR. Based on the results for the stationary scenarios, R2 and R3, our proposed system generally selects a better data rate and route based on more accurate link quality information and achieves higher end-to-end throughput.

\section{Conclusion}
\label{sec:Conclusion}

In this paper, we have proposed a new LQE method to overcome the deficiencies of the existing methods viz., accuracy, adaptability, environment awareness, ease of implementation and efficiency. We compared our proposed method with two general packet counting based methods in different scenarios. Based on a detailed comparison, the SNR profile based LQE has the advantage of high accuracy and low overhead, as shown via an analytical model. To overcome estimation inaccuracies due to environment changes, we proposed an updating mechanism for the SNR profile. Measurement results show the effectiveness of the profile updates, which suggest that the SNR profile based LQE can be a general LQE method used in all environments. Results have been presented indicating that our method generally selects a better rate and achieves higher throughput even in challenging, i.e. mobile, scenarios. Further, each of the procedures presented in the paper can be executed in polynomial time making it easily implementable.

We also used our proposed LQE method in the link quality aware data rate selection and routing mechanism. Our measurements indicate that our proposed system improves two aspects: (i) per-link throughput, and (ii) faster and better route selection. These improvements are mainly because of faster and better rate selection procedures and LQE.

The next task in our roadmap is to further reduce the error for LQE by combining both SNR profiling and packet counting methods. We intend to input the data from our measurement to a network simulation to verify further the performance of our methods. Through simulation we can test many scenarios and evaluate more metrics like scalability. Further in the roadmap, we shall extend our work to consider multiple packet sizes and improve the performance in rate selection to select better data rate. Also, self optimization work will be performed to let the system self-optimize the performance and decide the different $\alpha_x$ values.

\end{document}